\documentclass[%
reprint,
 amsmath,amssymb,
 aps,
prb,
superscriptaddress
]{revtex4-2}

\usepackage{graphicx}
\usepackage{dcolumn}
\usepackage{bm}
\usepackage{mathrsfs}
\usepackage{braket}
\usepackage{color}
\usepackage{comment}
\usepackage{siunitx}
\usepackage{makecell}
\usepackage{multirow}

\usepackage{hyperref}
\hypersetup{
     colorlinks   = true,
     linkcolor    = blue,
     citecolor    = blue,
     urlcolor     = blue
}

\begin{document}

\preprint{APS/123-QED}

\title{Riemannian geometric classification and emergent phenomena of magnetic textures}

\author{Koki Shinada}
\altaffiliation[koki.shinada@riken.jp]{}
\affiliation{
RIKEN Center for Emergent Matter Science (CEMS), Wako, Saitama 351-0198, Japan}
\author{Naoto Nagaosa}
\altaffiliation[nagaosa@riken.jp]{}
\affiliation{
RIKEN Center for Emergent Matter Science (CEMS), Wako, Saitama 351-0198, Japan}%
\affiliation{Fundamental Quantum Science Program (FQSP), TRIP Headquarters, RIKEN, Wako 351-0198, Japan}

\date{\today}

\begin{abstract}
We propose a new classification of magnetic textures from a differential-geometric viewpoint.
Magnetic textures are conventionally classified into collinear, coplanar, and noncoplanar magnets according to the relative arrangement of spins. 
These classes are typically characterized by the vector spin chirality (VSC) and the scalar spin chirality (SSC), which serve as indicators of noncollinearity and noncoplanarity, respectively.
However, this conventional classification is incomplete: in particular, noncoplanar textures cannot be fully characterized by the SSC alone, as exemplified by conical magnets.
To refine this classification, we analyze the curves and surfaces traced by spins in real space using differential geometry and introduce two novel scalar spin chiralities that properly characterize noncoplanarity: the geodesic scalar spin chirality and the torsional scalar spin chirality. 
These quantities are directly connected to the differential geometry of curves on surfaces: the former is associated with the geodesic curvature, which quantifies the deviation from a geodesic, while the latter is related to the torsion. 
Based on these chiralities, we identify three distinct classes of noncoplanar magnetic textures.
Furthermore, in close analogy with the roles of the VSC and the conventional SSC in emergent electrodynamics, the geodesic SSC gives rise to novel emergent phenomena. 
By constructing a semiclassical theory that incorporates nonadiabatic effects and higher-order gradients of the spatial modulation of spins, we demonstrate that the geodesic SSC induces an emergent band asymmetry, leading to a nonreciprocal response as a quantum geometric effect. This mechanism is a purely orbital effect, requiring no spin-orbit coupling, and the resulting discussion runs in parallel with the conventional picture of the topological Hall effect driven by the SSC.
The geometric viewpoint developed here will provide broad new insights into classification, quantum geometry, emergent electrodynamics, and a wider variety of emergent phenomena.
\end{abstract}

\maketitle

\section{Introduction}
Magnetism has long provided fertile ground for the emergence of rich physical phenomena, ranging from fundamental symmetry breaking to diverse functional responses in solids. 
Beyond conventional ferromagnets and antiferromagnets, recent advances have revealed a wide variety of nontrivial magnetic textures, including multiferroic magnets \cite{Tokura_2014,fiebig2016evolution,spaldin2019advances}, altermagnets \cite{PhysRevX.12.040501}, and large-scale magnetic textures such as skyrmions and beyond \cite{nagaosa2013topological,GOBEL20211,koraltan20262026skyrmionicsroadmap}.
These states are often accompanied by unconventional transport, optical, and electromagnetic phenomena, highlighting magnetic textures as a central platform for emergent phenomena in condensed matter physics \cite{RevModPhys.82.1539,tokura2018nonreciprocal}.

The classification of magnetic textures plays an essential role in organizing emergent phenomena and identifying their underlying mechanisms. 
Several complementary frameworks have been developed, including symmetry-based classifications \cite{Birss1964}, as well as topological classifications based on global geometric invariants \cite{zang2018topology}. 
While these approaches have been highly successful, they primarily focus on symmetry constraints or global geometry and do not directly and quantitatively capture the local geometric structure of spin configurations themselves.

From a local geometric viewpoint, magnetic textures are often classified as collinear, coplanar, or noncoplanar according to the relative arrangement of spins. 
Within this geometric classification scheme, the vector spin chirality (VSC) and the scalar spin chirality (SSC) have played central roles as quantitative indicators. 
The VSC characterizes the degree of noncollinearity between neighboring spins, 
while the SSC quantifies the solid angle formed by three spins and characterizes the noncoplanarity. 
Furthermore, they are deeply related to emergent electrodynamics \cite{GEVolovik_1987,Nagaosa_2012}, including magnetically induced electric polarization \cite{PhysRevLett.95.057205,PhysRevLett.96.067601,kimura2003magnetic}, the emergent magnetic field responsible for the topological Hall effect \cite{PhysRevLett.69.3593,PhysRevLett.83.3737,PhysRevLett.93.096806,PhysRevB.62.R6065,doi:10.1126/science.1058161,machida2010time,PhysRevLett.102.186602}, and the emergent electric-field effects \cite{PhysRevLett.107.136804,schulz2012emergent,PhysRevLett.98.246601,PhysRevLett.102.067201,hai2009electromotive,Nagaosa_2019}.
These quantities have therefore served as standard order parameters linking magnetic textures to emergent electrodynamics, as well as for classification.

Despite the long-standing use of the VSC and the SSC, this geometric classification encounters a fundamental limitation when applied to noncoplanar magnetic textures. 
In particular, the vanishing of the SSC does not necessarily imply a coplanar spin configuration. 
A notable example is provided by conical magnets, in which spins are genuinely noncoplanar while the SSC identically vanishes due to the effectively one-dimensional nature of the modulation. 
As a result, these magnets are indistinguishable from coplanar magnets within the conventional classification scheme, despite their clearly different geometric nature.

This ambiguity is not merely a matter of classification. 
Recent experiments have revealed nonreciprocal responses in noncoplanar magnetic systems \cite{doi:10.1126/sciadv.aat1115,PhysRevMaterials.8.044407,yokouchi2017electrical,PhysRevB.103.L220410,PhysRevB.103.184428,PhysRevLett.122.057206,doi:10.1073/pnas.2405839122,doi:10.1126/sciadv.adw8023,ltwf-zhj2,jiang2020electric,https://doi.org/10.1002/adma.202420614,doi:10.1126/sciadv.abd3703}, indicating that essential information relevant to emergent phenomena is not captured by the SSC alone. 
These observations raise a fundamental question: what is the minimal geometric information required to uniquely characterize noncoplanar magnetic textures and to determine their emergent responses? 
Addressing this question requires going beyond conventional chirality measures and calls for a framework that directly characterizes the intrinsic geometry of spin configurations.

In this work, we refine the classification of noncoplanar magnets by organizing the geometric structures formed by spin moments from the perspective of differential geometry. 
We consider classical spins with a fixed magnitude of unity. 
Depending on the dimensionality of their spatial modulation $d_\mathrm{p}$ (the number of independent parameters characterizing the real-space modulation of the spins), spins arranged in real space trace either curves ($d_\mathrm{p} = 1$) or surfaces ($d_\mathrm{p} \geq 2$) on the unit sphere.
As a result, the problem of classifying magnetic textures can be reduced to a problem in the differential geometry of curves and surfaces on the unit sphere. 
This viewpoint enables a systematic classification of magnetic textures based on geometric indicators that characterize these curves and surfaces.
Building on the conventional VSC and SSC, we introduce two new geometric indicators: the \textit{geodesic scalar spin chirality} and the \textit{torsional scalar spin chirality}. 
These quantities can be regarded as extensions of the SSC and, like the conventional SSC, become finite only in noncoplanar magnetic configurations. 
A crucial difference, however, is that while the conventional SSC requires at least two spatial dimensions ($d=2$) to be finite, the newly introduced scalar spin chiralities can take finite values even in one-dimensional systems with $d=1$ or $d_\mathrm{p} =1$.
Importantly, these new indicators possess clear interpretations in differential geometry. 
The geodesic SSC is determined by the geodesic curvature, which characterizes the deviation of a curve from a geodesic. 
In contrast, the torsional SSC is governed by the torsion of the curve. 
Because curves on the unit sphere are uniquely characterized by these two geometric quantities, the proposed framework provides a complete classification of one-dimensional noncoplanar magnets with $d=1$ or $d_\mathrm{p} =1$. 
According to the presence or absence of these indicators, noncoplanar magnetic textures are classified into three distinct classes. 
Within this classification, conical magnets, which are previously ambiguous in conventional schemes, are correctly identified as a distinct class of noncoplanar magnets.

Even more remarkably, the geodesic SSC describes new emergent phenomena, closely paralleling the conventional picture in which the SSC acts as an effective magnetic field (the real-space Berry curvature) on itinerant electrons and gives rise to the topological Hall effect \cite{nagaosa2013topological}. 
By incorporating nonadiabatic effects and higher-order gradients in the spatial modulation of spins into the semiclassical theory, we demonstrate that the geodesic SSC emerges in the semiclassical Hamiltonian and the equations of motion as a real-space quantum geometric effect.
Specifically, it manifests itself as an emergent band asymmetry and generates nonreciprocal transport without relying on spin-orbit coupling. 
The resulting nonreciprocal conductivity is therefore a purely orbital effect originating from magnetic textures.
Our results establish a general principle for nonreciprocal transport in spin-orbit-coupling-free magnetic systems, rooted in the geometry of spin textures.

The structure of this paper is as follows:
We discuss magnetic textures from the viewpoint of differential geometry and introduce the geodesic SSC and the torsional SSC in Sec.~\ref{sec_differential_geometry}.
In addition, we provide a new classification for collinear magnets, coplanar magnets, and noncoplanar magnets.
In Sec.~\ref{sec_semiclassical}, we construct the semiclassical theory, taking nonadiabatic effects into account, and we discuss an emergent response, the nonreciprocal response, induced by the effective band asymmetry, which is purely determined by the geodesic SSC.

\section{Differential geometry in magnetic textures} \label{sec_differential_geometry}
Magnetic textures can be classified into three categories according to the relative arrangement of their constituent spins: collinear magnets, coplanar magnets, and noncoplanar magnets.
In collinear magnets, all spins are aligned parallel or antiparallel to each other, whereas in coplanar magnets, spins lie within a single plane. 
Magnetic textures that belong to neither of these categories are referred to as noncoplanar magnets.

Conventionally, the vector spin chirality and the scalar spin chirality have been used as indicators for classifying these magnetic categories. 
However, we find that these quantities alone are insufficient, particularly for the classification of noncoplanar magnets. 
To address this limitation, we introduce new indicators, the \textit{geodesic scalar spin chirality} and the \textit{torsional scalar spin chirality}.
This classification based on these indicators is distinct from symmetry-based classifications and topological classifications. 
Instead, it constitutes a new classification that focuses on the local geometric structure of the curves and surfaces traced by spins in $d$-dimensional real space $\mathbb{R}^d$ ($d=1, 2, 3$) \cite{Kobayashi2019DGCS}.

For the following discussion, we introduce a parameter dimension $d_\mathrm{p} (=0,1,2)$.
We consider a spin $\bm{n}(\bm{x})$ parameterized by the real-space coordinate $\bm{x}\in\mathbb{R}^d$.
We assume that the spin $\bm{n}(\bm{x})$ has a fixed magnitude, $|\bm{n}(\bm{x})| = 1$, and we consider the continuous limit.
The spin vector always lies on the unit sphere $\mathbb{S}^2$.
When $\bm{n}(\bm{x})$ is independent of $\bm{x}$, we define the parameter dimension as $d_\mathrm{p}=0$, in which case $\bm{n}(\bm{x})$ is fixed to a single point on $\mathbb{S}^2$.
If $\bm{n}(\bm{x})$ depends only on a single parameter $\Theta(\bm{x})$ (for example, $\Theta(\bm{x})=\bm{q}\cdot\bm{x}$ as in a single-Q state), we set $d_\mathrm{p}=1$, in which case the spin draws a curve on $\mathbb{S}^2$.
Similarly, when $\bm{n}(\bm{x})$ depends only on two parameters $\bm{\Theta}(\bm{x}) = (\Theta_1(\bm{x}),\Theta_2(\bm{x}))$, we define $d_\mathrm{p}=2$, in which case the spin draws a surface on $\mathbb{S}^2$.
In three-dimensional real space, we can also consider spin textures depending on three parameters ($d_\mathrm{p}=3$) such as hopfions; however, such cases are not considered in this work.
By definition, $d_\mathrm{p} \leq d$.
Accordingly, we will discuss the geometry of curves and surfaces on $\mathbb{S}^2$ and examine the geometric meaning of each indicator.

\subsection{Vector spin chirality and scalar spin chirality}
\begin{figure*}[t]
\centering
\includegraphics[width=0.8\linewidth]{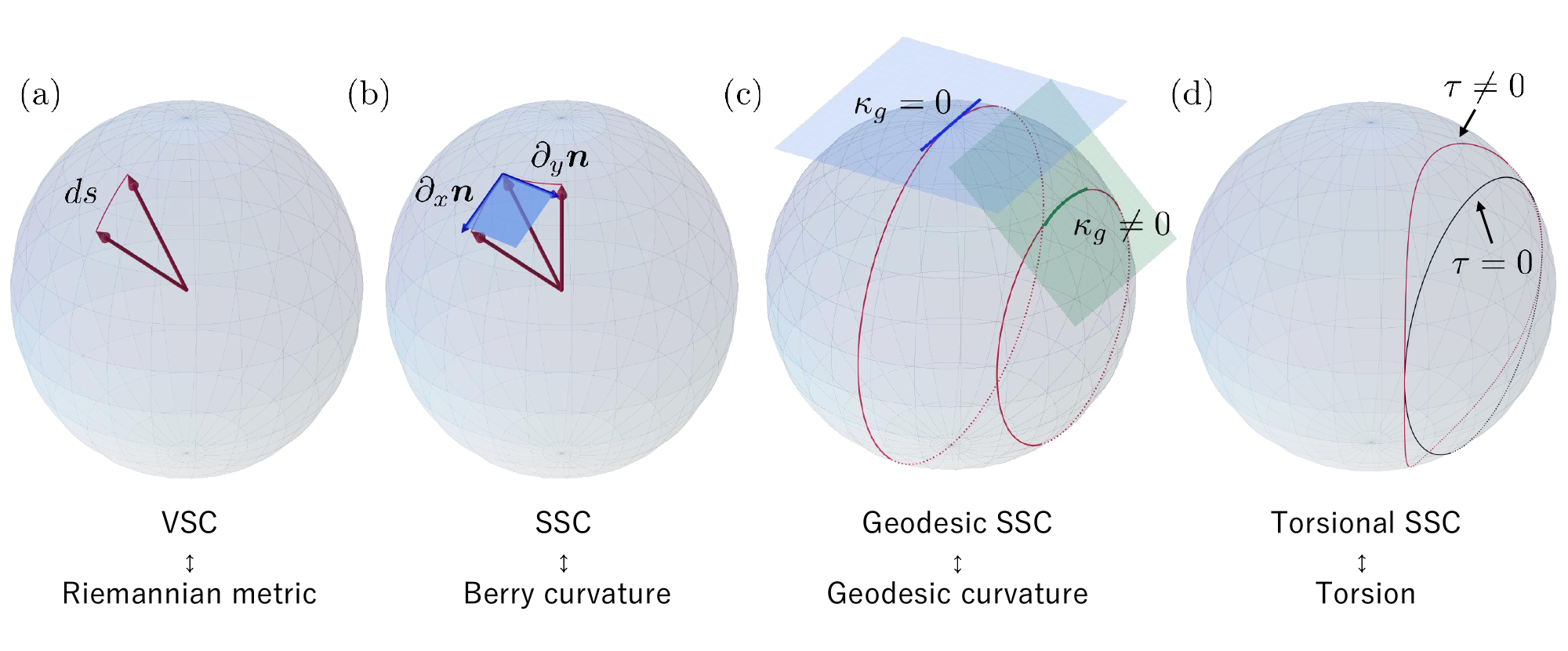}
\caption{
Schematic illustrations of the vector spin chirality (VSC), the scalar spin chirality (SSC), the geodesic SSC, and the torsional SSC in (a), (b), (c), and (d), respectively.
The geometric meanings of these quantities on $\mathbb{S}^2$ are as follows:
(a) VSC corresponds to the Riemannian metric $g_{ij}$ on $\mathbb{S}^2$, representing the infinitesimal distance $ds$ between nearby distinct points on $\mathbb{S}^2$ to which spins point, shown by the red segment.
(b) SSC corresponds to an oriented infinitesimal area spanned by two tangent vectors. The blue shaded region represents the area element.
(c) Geodesic SSC corresponds to the geodesic curvature $\kappa_g$.
A curve satisfying $\kappa_{g} = 0$ is called a geodesic and lies on a great circle on $\mathbb{S}^2$. 
The blue shaded plane is a tangent space at a point on a great circle and the projected curve onto the tangent space is straight, implying the intrinsic curvature of the curve is zero ($\kappa_g = 0$).
On the other hand, for a curve with finite $\kappa_g$, the projected curve is intrinsically curved, as shown in the green shaded tangent space.
(d) Torsional SSC corresponds to the torsion $\tau$.
A curve with $\tau = 0$ lies in a single plane and, when constrained on $\mathbb{S}^2$, is a circle.
If a curve deviates from a circle, it no longer lies in a single plane and acquires a finite torsion.
} \label{fig_geometry}
\end{figure*}
Here, we briefly review the vector spin chirality (VSC) and the scalar spin chirality (SSC), and discuss their geometric meanings and roles in classification.
The VSC $\bm{v}_i$ is defined by
\begin{equation}
    \bm{v}_i = \bm{n} \times \partial_i \bm{n}.
\end{equation}
Here, $\partial_i = \partial / \partial x_i$.
Whether this quantity vanishes or takes a finite value serves as an indicator of whether the spin $\bm{n}(\bm{x})$ is collinear or noncollinear.
In particular, the absolute value of the VSC is related to the Riemannian metric tensor $g_{ij}$ as
\begin{equation}
    |\bm{v}_i| = \sqrt{ \partial_i \bm{n} \cdot \partial_i \bm{n}} = \sqrt{g_{ii}}.
\end{equation}
The Riemannian metric $g$ characterizes the line element traced by the spin $\bm{n}(\bm{x}(t))$ on $\mathbb{S}^2$ when an arbitrary curve $\bm{x}(t)$ in $\mathbb{R}^d$ is specified by varying a single parameter $t$, as shown in Fig.~\ref{fig_geometry}(a).
Explicitly, 
the length of the curve $s(t)$ is given by
\begin{equation}
    s(t) 
    = 
    \int^t_0 \biggl| \frac{d \bm{n}}{dt'} \biggr| dt'
    =
    \int^t_0 \sqrt{ \partial_i \bm{n} \cdot \partial_j \bm{n} \frac{d x_i}{dt'} \frac{d x_j}{dt'} } dt',
\end{equation}
providing the infinitesimal distance $ds^2$ as
\begin{equation}
    ds^2 = \partial_i \bm{n} \cdot \partial_j \bm{n} dx_i dx_j \equiv g_{ij} dx_i dx_j.
\end{equation}
Unless otherwise stated, we use the Einstein summation convention.
When $g_{ij}$ vanishes, no curve is traced on $\mathbb{S}^2$, and the spin is confined to a single point, meaning the parameter dimension is zero ($d_\mathrm{p} =0$).
In this case, spins are collinear. 
By contrast, when $g_{ij}$ is finite, a curve or a surface is traced on $\mathbb{S}^2$, and the magnetic configuration is no longer collinear.

The SSC is defined by
\begin{equation}
    \chi_{ij} = \bm{n} \cdot (\partial_i \bm{n} \times \partial_j \bm{n}).
\end{equation}
This corresponds to the Berry curvature, namely, an oriented infinitesimal area element spanned by two tangent vectors $\partial_i \bm{n}, \partial_j \bm{n}$ on $\mathbb{S}^2$, as shown in Fig.~\ref{fig_geometry}(b).
In particular, for systems with $d=2$, the SSC yields a topological number that counts how many times the mapping from $(x,y)$ to $\bm{n}(x,y)$ wraps $\mathbb{S}^2 $.
This quantity is called the skyrmion number and is defined as
\begin{equation}
    N_{\mathrm{sk}} = \frac{1}{4\pi} \int_{\mathbb{R}^2} \chi_{xy} dx dy \in \mathbb{Z}.
\end{equation}
By definition of the SSC, a finite value of $\chi_{ij}$ requires three vectors $\bm{n}, \partial_i \bm{n}, \partial_j \bm{n}$ that have mutually orthogonal components, implying that the spins are not arranged coplanarly and are therefore noncoplanar.
In other words, when the spins are collinear or coplanar, $\chi_{ij}$ vanishes.
However, it should be noted that a vanishing $\chi_{ij}$ does not necessarily imply a coplanar configuration, whereas the converse statement holds.
Thus, noncoplanar magnetic configurations can exist even when $\chi_{ij}$ vanishes. 
Indeed, $\chi_{ij}$ is always zero in systems with $d=1$ or $d_\mathrm{p} =1$, yet noncoplanar magnetic states, such as conical magnets, do exist.
Therefore, it is clear that $\chi_{ij}$ alone is insufficient for classifying noncoplanar magnets.

\subsection{Geodesic scalar spin chirality}
We introduce the following quantity as a new indicator of noncoplanar magnets:
\begin{equation}
    \gamma_{ijk} = \bm{n} \cdot ( \partial_i \bm{n} \times \partial_{jk} \bm{n}  ).
\end{equation}
Here, $\partial_{jk} = \partial^2/\partial x_j \partial x_k$.
Trivially, $\gamma_{ijk} = \gamma_{ikj}$.
As in the case of the SSC, $\gamma_{ijk}$ represents a type of scalar chirality, which forms a scalar triple product, and requires three vectors $\bm{n}, \partial_i \bm{n}, \partial_{jk}\bm{n}$ that have mutually orthogonal components.
It therefore vanishes for collinear and coplanar magnets.
A crucial difference from the SSC, however, is that $\gamma_{ijk}$ remains finite even in systems with $d=1$ or $d_\mathrm{p} = 1$.
In the following, we discuss the geometric interpretation of $\gamma_{ijk}$.

\subsubsection{Geodesic curvature} \label{sec_geodesic_curvature}
Here, we consider situations in which the spin $\bm{n}(\bm{x})$ traces a curve on $\mathbb{S}^2$.
This includes not only the cases of $d=1$ ($\bm{n}(x)$) or $d_\mathrm{p}=1$ ($\bm{n}(\Theta(\bm{x}))$), but also higher-dimensional systems with $d=2,3$, provided that a curve $\bm{x}(t)$ in real space $\mathbb{R}^d$ is specified ($\bm{n}(\bm{x}(t))$).
We discuss the differential geometry of curves $\bm{n}(w)$ on $\mathbb{S}^2$ parameterized by a general single parameter $w \in \mathbb{R}$. $w$ can be $x$, $\Theta$, or $t$.

We introduce the arc-length parameter as
\begin{equation}
    s = \int^w_0 | \partial_{w'} \bm{n}(w') | dw',~~~ds = |\partial_w \bm{n}(w)| dw.
\end{equation}
If there is no point at which $|\partial_w \bm{n}(w)| = 0$ along the path, the arc-length parameter $s$ is a monotonically increasing function of $w$. In this case, a one-to-one correspondence exists between $w$ and $s$, and the parameter $w$ can be reparameterized in terms of $s$.
When the parameter is chosen to be $s$, the magnitude of the spin variation with respect to $s$ (i.e. the modulation speed) is always unity ($|\bm{n}'(s)| = 1$), which simplifies the following analysis.
In the following, a prime denotes differentiation with respect to $s$.
On $\mathbb{S}^2$, $\bm{n}$ is a unit normal vector, while $\bm{n}'$ is a unit tangent vector along the curve.
In general, $\bm{n}''$ has both tangential and normal components, however, it is orthogonal to $\bm{n}'$ because of $|\bm{n}'| = 1$.
This vector is explicitly given by
\begin{equation} \label{geodesic}
    \bm{n}'' = - \bm{n} + \kappa_g (\bm{n} \times \bm{n}').
\end{equation}
The first term represents the normal component and originates from the curvature of the unit sphere itself.
This coefficient $-1$ follows from the identity $\bm{n} \cdot \bm{n}'' = - \bm{n}' \cdot \bm{n}' = -1$, using $|\bm{n}'|=1$.
However, this normal component is not essential for the following discussion.
The second term is proportional to the other unit tangent vector that is orthogonal to $\bm{n}'$, and its proportionality coefficient $\kappa_g(s) (= \bm{n}'' \cdot ( \bm{n} \times \bm{n}') )$ is referred to as the geodesic curvature.
The geodesic curvature $\kappa_g$ measures the curvature of the curve itself and can be finite even for a curve lying on a flat plane.
Furthermore, a curve that satisfies $\kappa_g = 0$ at every point along it is called a geodesic.
This indicates that when a geodesic is projected onto the tangent plane, it becomes a straight line. 
In contrast, a curve with a finite geodesic curvature $\kappa_g$ is intrinsically curved, and its projection onto the tangent plane is also curved, as shown in Fig.~\ref{fig_geometry}(c).
On $\mathbb{S}^2$, geodesics correspond to curves lying on great circles.
The sign of the geodesic curvature is determined by whether $\bm{n}''$ is parallel (plus sign) or antiparallel (minus sign) to $\bm{n} \times \bm{n}'$ on the tangent space.

\begin{figure}[t]
\centering
\includegraphics[width=1.0 \linewidth]{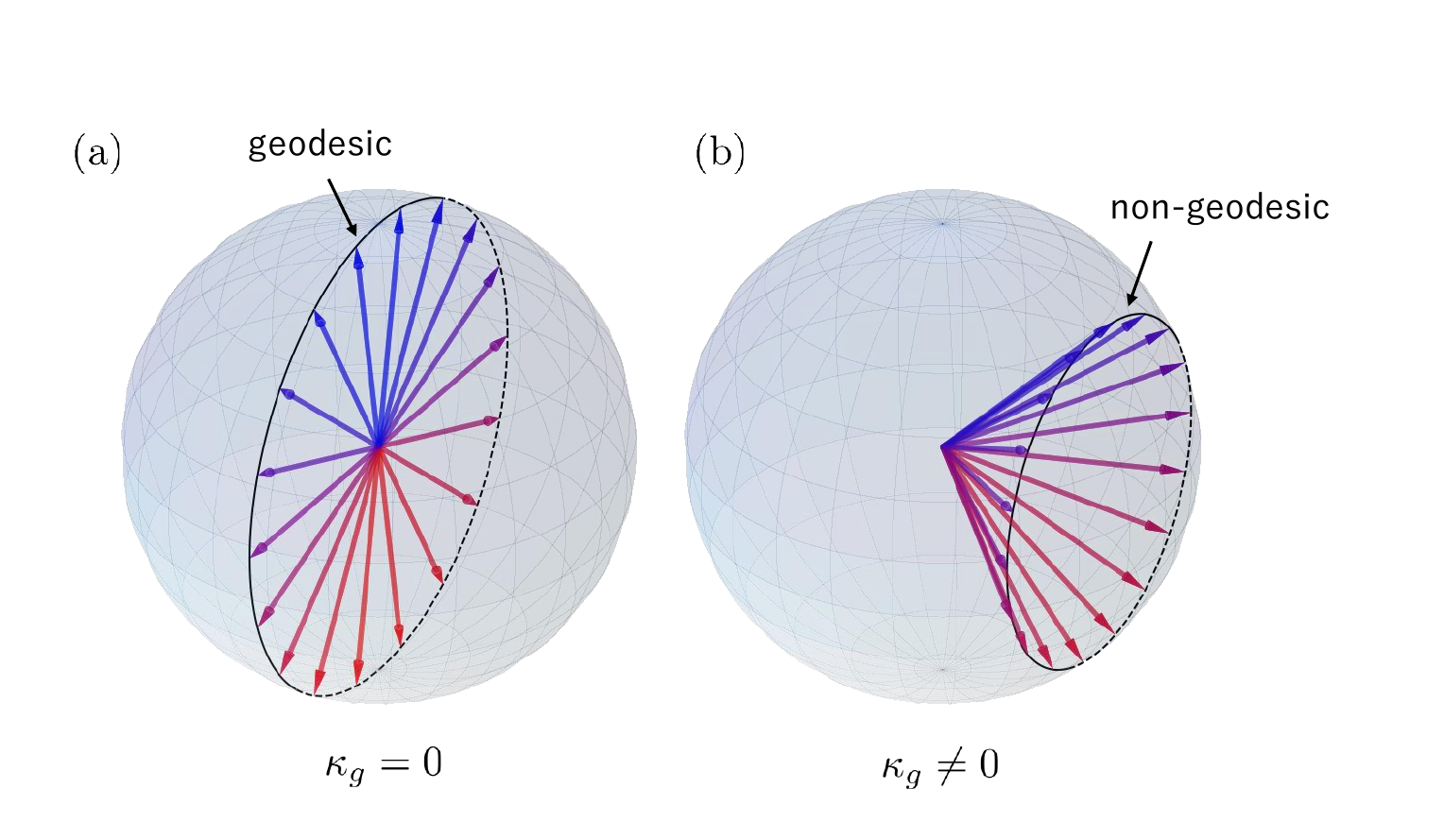}
\caption{
Mapping of helical and conical magnetic textures to the unit sphere $\mathbb{S}^2$.
For helical magnets, the spin trajectories trace geodesics (great circles) on the unit sphere, and hence the geodesic curvature $\kappa_g$ vanishes. 
In contrast, the trajectories for conical magnets are not geodesics; therefore their geodesic curvature $\kappa_g$ is finite.
} \label{fig_geodesic_magnet}
\end{figure}

We now obtain the relation between $\gamma_{www}$ and the geodesic curvature $\kappa_g$, using Eq.~(\ref{geodesic}), as
\begin{equation}
    \gamma_{www} = \kappa_g | \partial_w \bm{n}|^3,
\end{equation}
where we use the relation $ds/dw = |\partial_w \bm{n}|$.
Therefore, this expression shows that coplanar magnets, such as helical magnets and chiral soliton lattices in which the spin traces a curve on a great circle in Fig.~\ref{fig_geodesic_magnet}(a), always yield zero $\gamma_{www
}$.
Thus, $\gamma_{www}$ is finite exclusively in noncoplanar magnets.
Since $\gamma_{www}$ is related to the geodesic curvature, we refer to it here as the geodesic scalar spin chirality (geodesic SSC).

Representative examples that yield a finite value of $\gamma_{www}$ are conical magnets. 
The spin structure of conical magnets is given by
\begin{equation} \label{eq_conical}
    \bm{n}(\bm{x}) = ( \cos \alpha \cos qx , \cos \alpha \sin qx , \sin \alpha  ),
\end{equation}
and the mapping to $\mathbb{S}^2$ is shown in Fig.~\ref{fig_geodesic_magnet}(b).
Here, $\alpha$ is the spin canting angle and is constant in space, $q$ is a wavenumber, and $w = \Theta(\bm{x}) = qx$.
In the case of $\alpha=0$, $\bm{n}(\bm{x})$ is reduced to the helical magnetic structure in Fig.~\ref{fig_geodesic_magnet}(a).
In the case of conical magnets, the spin trajectory traces a small circle and is therefore not a geodesic, resulting in a finite geodesic curvature $\kappa_g$.
The geodesic SSC reads
\begin{equation} \label{eq_gamma_conical}
    \gamma_{xxx} = q^3 \cos^2 \alpha \sin \alpha,
\end{equation}
which is constant in space. In the case of helical magnets ($\alpha=0$), it vanishes.
If the sign of $\alpha$ or $q$ is inverted, where the corresponding spin configuration is generated by applying time reversal or spatial inversion to the original one, the sign of $\gamma_{xxx}$ is also inverted.
This can also be confirmed from the fact that reversing the sign of $\alpha$ or $q$ interchanges the parallel and antiparallel alignment between $\bm{n}''$ and $\bm{n} \times \bm{n}'$ on the tangent space.

\subsubsection{Gauss-Bonnet theorem}
Here, we consider the case in which the curve traced by $\bm{n}(w)$ is closed or periodic.
We denote the period by $L$, which can be infinite when the spin orientations coincide at $w = -\infty$ and $w = \infty$.
In the above discussion, whether the curve is open or closed is not essential.
For a closed or periodic curve, we show that, under a certain condition, the integral of $\gamma_{www}$ over one period can be evaluated straightforwardly using the Gauss-Bonnet theorem.

For a smooth closed curve $\bm{n}(s)$ parametrized by the arc-length parameter, the Gauss-Bonnet theorem states the relation
\begin{equation}
    \int_\Sigma K dA + \oint_{\partial \Sigma} \kappa_g ds = 2\pi.
\end{equation}
Here, $\Sigma$ is a two-dimensional subsurface, and $\partial \Sigma$ is its oriented boundary.
The orientation is chosen such that $\Sigma$ lies on the left-hand side when moving along the curve $\partial \Sigma$.
$K$ is the Gaussian curvature.
In particular, on $\mathbb{S}^2$, the Gaussian curvature is constant and equal to unity.
Then, we obtain the relation in the case of $\mathbb{S}^2$
\begin{equation}
    \oint_{\partial \Sigma} \kappa_g ds = 2\pi - \mathrm{Area}(\Sigma).
\end{equation}
Here, $\mathrm{Area}(\Sigma)$ is the area of the surface $\Sigma$.

On the other hand, the integral of the geodesic SSC over one period is given by
\begin{equation}
    \braket{\braket{\gamma_{www}}} = \oint_0^{L} \kappa_g |\partial_w \bm{n}|^3 dw
    =
    \oint_{\partial \Sigma} \kappa_g |\partial_w \bm{n}|^2 ds.
\end{equation}
In the following, we assume $|\partial_w \bm{n}| = \mathrm{const} \equiv |q_0|$.
This condition implies that the spin traces the curve at constant speed.
This assumption is satisfied in helical and conical magnets.
In this case, using the Gauss-Bonnet theorem, we obtain
\begin{equation} \label{eq_gauss_bonnet_area}
    \braket{\braket{\gamma_{www}}} = |q_0|^2 (2\pi - \mathrm{Area}(\Sigma)).
\end{equation}
Therefore, when the speed is constant, the geodesic SSC is determined solely by the speed and the area enclosed by the curve. 

We apply Eq.~(\ref{eq_gauss_bonnet_area}) to simple cases.
For a helical magnet, the spin trajectory traces a great circle on $\mathbb{S}^2$.
The region enclosed by the great circle corresponds to a hemisphere of area $2\pi$.
As a result, the spatial average vanishes.
In contrast, for a conical magnet, considering the case $q>0$ in Eq.~(\ref{eq_conical}), the small circle encloses an area $2\pi(1 - \sin \alpha)$ on $\mathbb{S}^2$.
The speed of the modulation is constant with $|\partial_x \bm{n}| = q |\cos \alpha|$.
Using Eq.~(\ref{eq_gauss_bonnet_area}), we obtain $\braket{\braket{\gamma_{xxx}}} = 2\pi q^2 \cos^2 \alpha \sin \alpha$.
We can directly evaluate the integral of $\gamma_{xxx}$ over one period using Eq.~(\ref{eq_gamma_conical}) and confirm that the same result is obtained when the period $L = 2\pi/q$ is taken into account.
The analysis for $q<0$ proceeds analogously, except that the enclosed area on $\mathbb{S}^2$ acquires the opposite orientation.

\subsubsection{Two-dimensional case} \label{sec_two_dim}
We consider two-dimensional systems where the spin $\bm{n}(\bm{x})$ is parametrized by two general variables, $\bm{w}=(u,v) \in \mathbb{R}^2$.
This situation corresponds to $\bm{w} = (x,y)$ in the case of $d=2$, and to $\bm{w}=(\Theta_1(\bm{x}),\Theta_2(\bm{x}))$ when $d_\mathrm{p}=2$.
It also encompasses the case of a subsurface in $\mathbb{R}^3$ $\bm{x}(t_1,t_2)$ defined over a two-parameter domain $(t_1,t_2)$.
The tangent space on $\mathbb{S}^2$ is spanned by two linearly independent basis vectors, $\partial_u \bm{n}$ and $\partial_v \bm{n}$.
Since the normal vector is given by $\bm{n}$ itself on $\mathbb{S}^2$, any vector in $\mathbb{R}^3$ can be expressed as a linear combination of $\{ \bm{n}, \partial_u \bm{n}, \partial_v \bm{n} \}$.
The second derivatives of $\bm{n}$ are given by the following relation:
\begin{equation}
    \partial^w_{ij} \bm{n} = \Gamma^k_{ij} \partial^w_{k} \bm{n} + L_{ij} \bm{n}.
\end{equation}
Here, $\Gamma^k_{ij}$ are the Christoffel symbols, and $L_{ij}$ are the coefficients of the second fundamental form.
The superscript $w$ in symbols such as $\partial_i^{w}$ denotes partial derivatives with respect to the parameter $\bm{w}$.
The second derivatives of $\bm{n}$ describe the variation of the tangent vectors.
The first term, involving the Christoffel symbols, represents the variation within the tangent space, while the second term corresponds to the variation in the direction normal to the tangent space.
Using this relation, $\gamma^w_{ijk}$ is rewritten by
\begin{equation}
    \gamma^w_{ijk} = \Gamma_{jk}^l \chi^w_{il}.
\end{equation}
Since there exists no coordinate system on the sphere in which the Christoffel symbols vanish everywhere, a finite SSC generally produces a finite $\gamma^w_{ijk}$.

The components $\gamma_{uuu}$ and $\gamma_{vvv}$ can be understood in terms of the geodesic curvature, in the same manner as discussed in Sec.~\ref{sec_geodesic_curvature}.
In contrast, mixed components such as $\gamma_{uuv}$ are specific to two-dimensional systems.
They are related to the dipole component of the spatial modulation of the SSC.
Indeed, the derivative of the SSC is directly related to $\gamma^w_{ijk}$ as
\begin{equation}
    \partial^w_{i} \chi^w_{jk} = \gamma^w_{jik} - \gamma^w_{kij},
\end{equation}
where the right-hand side is composed solely of mixed components, because each of the components $\gamma_{uuu}$ and $\gamma_{vvv}$ cancels itself within this expression and provides no nontrivial relation. Note here that $\partial_i \bm{n} \cdot ( \partial_j \bm{n} \times \partial_k \bm{n} ) = 0$ because there are no three linearly independent vectors perpendicular to $\bm{n}$.

\begin{figure}[t]
\centering
\includegraphics[width=1.0 \linewidth]{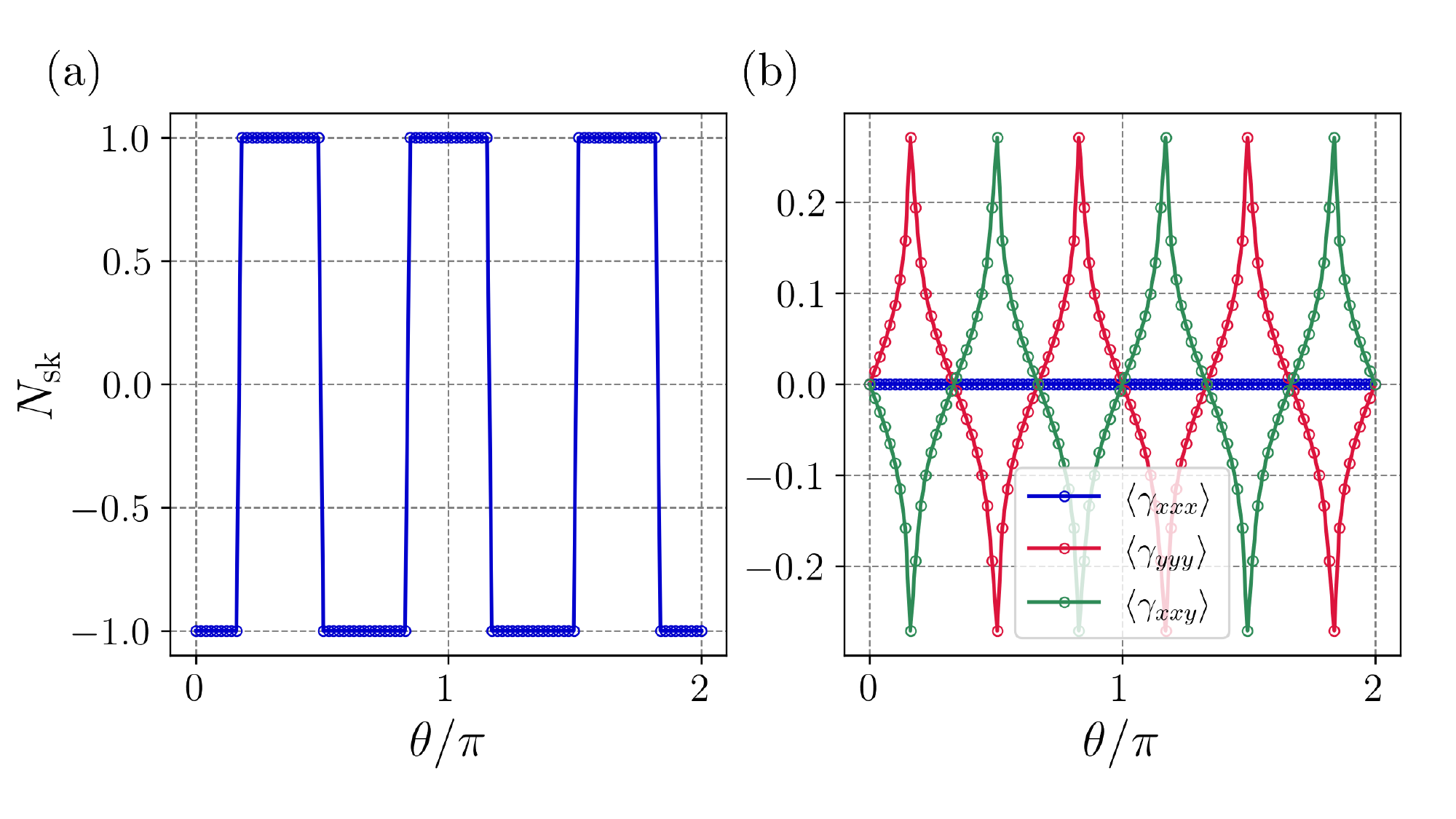}
\caption{(a) $\theta$-dependence of the skyrmion number $N_{\mathrm{sk}}$.
(b) $\theta$ dependence of the spatial average of $\gamma_{xxx}$ (blue), $\gamma_{yyy}$ (red), and $\gamma_{xxy}$ (green) in the unit cell. We set $\braket{\gamma_{ijk}}$ in units of $Q^3$.
} \label{fig_2d_nsk_gamma}
\end{figure}
\begin{figure*}[t]
\centering
\includegraphics[width=1.0 \linewidth]{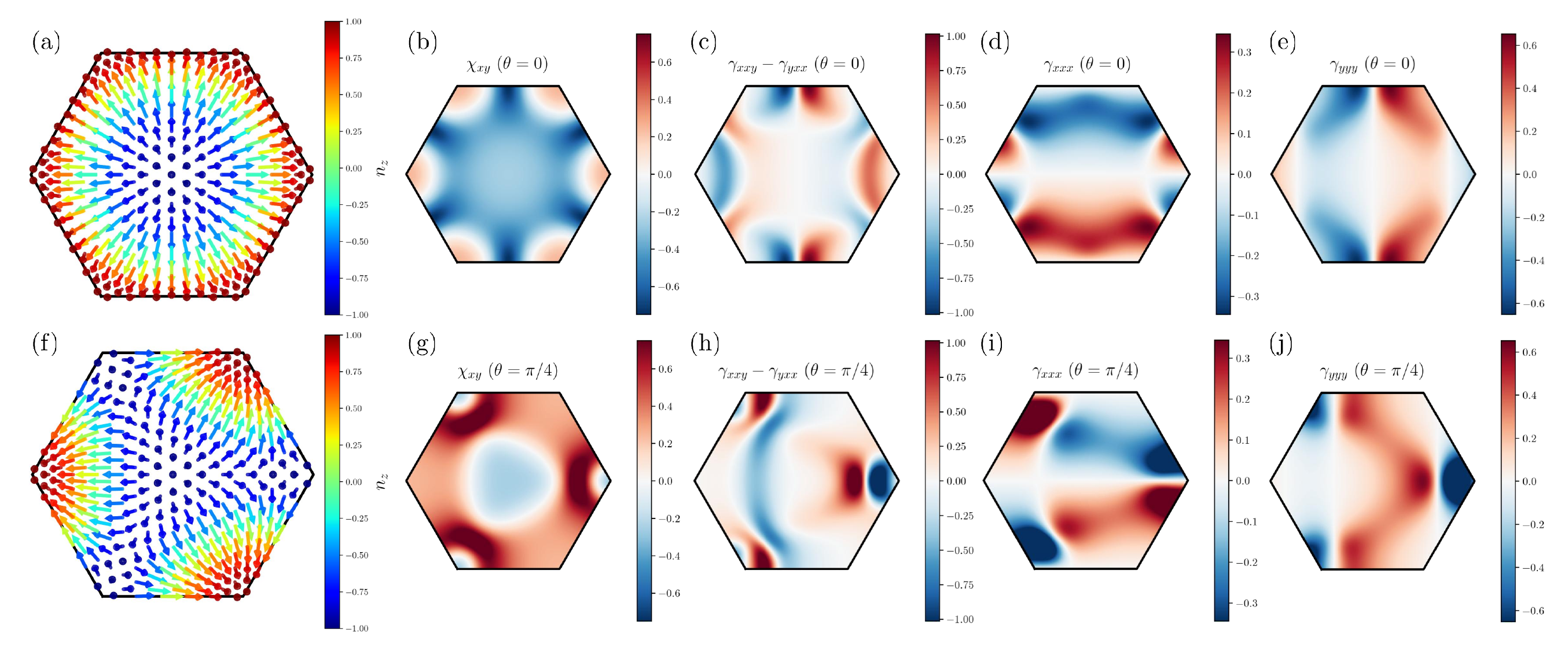}
\caption{
(a), (f) Spin textures described by the normalized spin $\bm{n}(\bm{x}) = \bm{N}(\bm{x})/|\bm{N}(\bm{x})|$ in Eq.~(\ref{eq_2d_skyrmion}) within a unit cell at $\theta =0 $ and $\theta = \pi/4$.
Spatial distributions of $\chi_{xy}$ [(b), (g)], $\gamma_{xxy} - \gamma_{yxx}$ [(c), (h)], $\gamma_{xxx}$ [(d), (i)], and $\gamma_{yyy}$ [(e), (j)].
We set $\chi_{xy}$ and $\gamma_{ijk}$ in units of $Q^2$ and $Q^3$, respectively.
} \label{fig_2d_all_figure}
\end{figure*}

We examine the behavior of $\gamma^w_{ijk}$ for skyrmion crystals in $\mathbb{R}^2$, and we set $\bm{w}=(x,y)$.
We use triple-Q spin textures described by an unnormalized vector \cite{hayami2021phase,PhysRevB.110.L241108}
\begin{equation} \label{eq_2d_skyrmion}
    \bm{N}(\bm{x}) = \sum_{\nu=1}^3 ( \sin Q_\nu \cos \phi_\nu , \sin Q_\nu \sin \phi_\nu, - \cos Q_\nu ).
\end{equation}
In the following, we use the normalized vector $\bm{n} = \bm{N}/|\bm{N}|$.
Using the three Q vectors defined as $\bm{Q}_1=Q(1,0)$, $\bm{Q}_2 = Q(-1/2 , \sqrt{3}/2)$, and $\bm{Q}_3 = Q(-1/2 , -\sqrt{3}/2)$, we define $Q_\nu = \bm{Q}_\nu \cdot \bm{x} + \theta$.
We also define $\phi_\nu = 2\pi (\nu -1)/3$.
The $\theta$ dependence of the skyrmion number $N_{\mathrm{sk}}$ is shown in Fig.~\ref{fig_2d_nsk_gamma} (a).
The skyrmion number exhibits sign reversals at $\theta = (2n+1)\pi/6$ ($n$ are integers).
At these points, the magnitude of $\bm{N}$ is zero at a certain point, rendering the spin $\bm{n}$ ill-defined.
Such points will be excluded from the following discussion.
At $\theta = 0$, the system realizes a N\'eel-type skyrmion crystal corresponding to a skyrmion with vorticity $m=1$ and helicity $\gamma=0$, as shown in Fig.~\ref{fig_2d_all_figure}(a).

Figure \ref{fig_2d_nsk_gamma}(b) displays the $\theta$ dependence of the spatial average of $\gamma_{xxx}$, $\gamma_{yyy}$, and $\gamma_{xxy}$ in the unit cell, which are defined by $\braket{\gamma_{ijk}} = \int_{\mathrm{u.c.}} d^2 \bm{x} \gamma_{ijk} / A_{\mathrm{u.c.}}$, where $A_{\mathrm{u.c.}}$ is the area of the unit cell.
Since the system always preserves $C_{3z}$ and $m_y \mathcal{T}$ symmetries, the relations $\braket{\gamma_{xxx}} = \braket{\gamma_{xyy}} = \braket{\gamma_{yxy}}=0$ and $\braket{\gamma_{yyy}} = - \braket{\gamma_{xxy}} = - \braket{\gamma_{yxx}}$ hold, as can be verified in this figure.
In the case of $\theta = n\pi/3$, the system has the $C_{2z}$ symmetry, and all components of $\braket{\gamma_{ijk}}$ vanish.
Furthermore, near $\theta = (2n+1)\pi/6$, where the skyrmion number changes sign, components such as $\gamma_{yyy}$ exhibit enhanced behavior. We summarize the presence or absence of the spatial average of $\chi_{xy}$ and $\gamma_{ijk}$ in Table~\ref{table_ad_chi_gamma}.

Figure \ref{fig_2d_all_figure} shows the spatial distributions of the SSC and the geodesic SSC within the unit cell for $\theta = 0$ ($N_{\mathrm{sk}} = -1$) and $\theta = \pi/4$ ($N_{\mathrm{sk}} = +1$).
The corresponding spin textures are shown in panels (a) and (f).
The spatial profiles of the SSC are shown in panels (b) and (g), respectively.
As discussed above, the mixed $x$-$y$ components of $\gamma_{ijk}$ are related to the dipolar structure of the SSC.
Here, we focus on $\partial_x \chi_{xy} = \gamma_{xxy} - \gamma_{yxx}$, which is shown in panels (c) and (h).
As is evident from the comparison with the spatial distribution of the SSC in panels (b) and (g), panels (c) and (h) represent the dipole structures of the SSC along the $x$ direction.
The spatial distributions of $\gamma_{xxx}$ and $\gamma_{yyy}$ are also plotted in panels (d), (i) and (e), (j), respectively.
In two dimensions, $\gamma_{xxx}$ can be interpreted as the collection of geodesic curvatures of the curves traced by the spin $\bm{n}$ on $\mathbb{S}^2$ when sweeping along the $x$ direction with $y$ fixed.
An analogous interpretation applies to $\gamma_{yyy}$, corresponding to sweeps along the $y$ direction with $x$ fixed.
Some regions where they vanish (shown in white) originate from the fact that, at points in the regions, the spin texture locally follows geodesics, namely trajectories along great circles on $\mathbb{S}^2$.
For example, inspection of the spin structure along the white lines extending in the $x$ direction in panels (d) and (i) reveals that the spins trace a great circle in a manner analogous to helical magnets as can be seen in Fig.~\ref{fig_2d_all_figure}(a), (f).
Because the system preserves the $m_y\mathcal{T}$ symmetry, $\chi_{xy}$ and $\gamma_{yyy}$ take the same values at points related by a mirror reflection perpendicular to the $y$ axis.
In contrast, $\gamma_{xxx}$ changes sign while retaining the same magnitude.
Consequently, the unit-cell integral of $\gamma_{xxx}$ always vanishes, as discussed above.
At $\theta = 0$, the system additionally preserves the $m_x\mathcal{T}$ symmetry.
Accordingly, at points related by a mirror reflection perpendicular to the $x$ axis, $\chi_{xy}$ and $\gamma_{xxx}$ take the same values, whereas $\gamma_{yyy}$ changes sign while retaining the same magnitude.
As a result, the unit-cell integral of $\gamma_{yyy}$ also vanishes, as discussed above.
In contrast, at $\theta = \pi/4$ such symmetries are absent, and $\gamma_{yyy}$ does not cancel between positive and negative contributions, as confirmed in Fig.~\ref{fig_2d_all_figure}(j).

\begin{table}
\centering
\caption{Summary of the presence (\checkmark) or absence (zero) of the spatial averages $\chi_{xy}$ and $\gamma_{ijk}$ in triple-Q spin textures described by Eq.~(\ref{eq_2d_skyrmion}).}  \label{table_ad_chi_gamma} 
\renewcommand{\arraystretch}{1.3}
\begin{tabular}{@{\hskip 5pt}c@{\hskip 5pt}||@{\hskip 20pt}c@{\hskip 20pt}@{\hskip 20pt}c@{\hskip 20pt}} \hline
      $\theta$ value & $\theta = n \pi/3$  & otherwise \\
    \hline 
     $\braket{\chi_{xy}}$    &    \checkmark  & \checkmark \\
     $\braket{\gamma_{ijk}}$    &    zero  & \checkmark \\
     Symmetry & $m_x\mathcal{T},m_y\mathcal{T}$ & $m_y \mathcal{T}$ \\
     \hline
\end{tabular}
\end{table}

\subsubsection{Symmetry argument}
For the spatial integral of $\gamma_{ijk}$ to be finite, spatial inversion symmetry must be broken.
In the presence of spatial inversion symmetry, $\bm{n}( - \bm{x}) = \bm{n}(\bm{x})$ holds.
Since $\gamma_{ijk}$ involves three spatial derivatives, the spatial integral is forced to vanish by this symmetry.

Moreover, in magnetic systems without spatial inversion symmetry, the local $\gamma_{ijk}$ and its spatial integral change sign between spin textures, $\bm{n}(\bm{x})$ and $\bm{n}_{\mathcal{P}}(\bm{x})$, that are related to each other by spatial inversion.
In particular, for a one-dimensional system with $d=1$ or $d_\mathrm{p} = 1$, spatial inversion corresponds to reversing the orientation of the curve traced by the spin on $\mathbb{S}^2$.
As a consequence, the geodesic curvature $\kappa_g$ and the orientation of the area enclosed by the curve are also reversed.
Namely, spatial inversion maps the originally enclosed surface $\Sigma$ onto $\Sigma_{\mathcal{P}}$, and these areas satisfy the relation $\mathrm{Area}(\Sigma_{\mathcal{P}}) = 4\pi - \mathrm{Area}(\Sigma)$.
Using this relation, in the case of the constant spin modulation in Eq.~(\ref{eq_gauss_bonnet_area}), the spatial integral of the spatially inverted $\gamma_{xxx}$ is given by
\begin{equation}
    \braket{\braket{\gamma_{xxx}^{\mathcal{P}}}} = |q_0^2| ( 2\pi - \mathrm{Area}(\Sigma_{\mathcal{P}}) )
    = - \braket{\braket{\gamma_{xxx}}},
\end{equation}
showing the sign inversion.
Furthermore, both the local $\gamma_{ijk}$ and its spatial integral change sign between spin textures ($\bm{n}(\bm{x})$ and $\bm{n}_{\mathcal{T}}(\bm{x})$) that are related by time-reversal operation.

In the multipole classification, $\gamma_{ijk}$ is a third-rank tensor that is odd under both time reversal and spatial inversion symmetries, and thus corresponds to a magnetic toroidal octupole.
Since this is a reducible tensor, its irreducible decomposition contains a magnetic toroidal dipole moment, a magnetic quadrupole, and a magnetic toroidal octupole \cite{PhysRevB.104.054412}.
While the conventional scalar spin chirality represents an emergent magnetic dipole field density, the geodesic scalar spin chirality can instead be regarded as the density of such multipoles characterizing magnetic structures that break spatial inversion symmetry.
Applying this framework to hopfions and other magnetic textures with magnetic toroidal moments \cite{PhysRevResearch.2.013315,PhysRevB.104.075102,PhysRevLett.129.267201} may provide an interesting future direction.

\subsection{Torsional scalar spin chirality}
Here, we consider one-dimensional systems described by a single parameter $w \in \mathbb{R}$ with $d=1$ or $d_\mathrm{p} = 1$.
We introduce the following quantity, which becomes finite only in noncoplanar magnets:
\begin{equation}
    t = \partial_w \bm{n} \cdot ( \partial_{ww} \bm{n} \times \partial_{www} \bm{n} ). 
\end{equation}
This quantity is related to the torsion $\tau$.
We refer to $t$ as the torsional scalar spin chirality.
For a curve on $\mathbb{S}^2$, we obtain (see Appendix \ref{appendix_torsion} for the detailed derivation)
\begin{equation}
    t = \tau (\kappa_g^2 + 1) | \partial_w \bm{n} |^6.
\end{equation}
In general, a curve satisfying $\tau = 0$ lies in a plane, whereas a finite torsion implies that the curve is not in a single plane, as shown in Fig.~\ref{fig_geometry}(d).
For example, the helical and conical magnetic states trace great and small circles on $\mathbb{S}^2$, respectively, and the curves lie in a single plane.
In general, among curves on $\mathbb{S}^2$, those lying in a single plane are restricted to circles.
Indeed, cutting a sphere with a plane yields a circle as the intersection, which directly confirms this statement.
In this case, $\tau$ and $t$ vanish; otherwise, they can be finite.
We note that the planarity of the spin trajectory is different from the coplanarity of the spins: the former means that the curve on $\mathbb{S}^2$ lies in a plane, whereas the latter means that all spins in real space lie in a single plane.
As in conical magnets, even in noncoplanar magnets, the curve traced by the spin can lie within a single plane.

Explicitly, the torsion $\tau$ of a curve on $\mathbb{S}^2$ is related to the geodesic curvature $\kappa_g$ through $\tau = \kappa_g'/(1 + \kappa_g^2)$. Thus, if $\kappa_g$ is constant, the torsion is automatically zero.
In particular, great and small circles have a constant $\kappa_g$, and hence $\tau=0$.
As an example that yields a finite torsion, we consider the spin texture defined by an unnormalized vector
\begin{equation} \label{eq_dist_conical}
    \bm{N}(\bm{x}) = ( \cos \alpha \cos qx , 2 \cos \alpha \sin qx , \sin \alpha ).
\end{equation}
This texture lacks the equivalence between the $x$ and $y$ directions of the spin, in contrast to conical or helical magnets.

\begin{table*}[t]
  \renewcommand{\arraystretch}{1.3}
  \centering
  \caption{Classification of magnetic textures based on the vector spin chirality $\bm{v}_i$, the scalar spin chirality (SSC) $\chi_{ij}$, the geodesic SSC $\gamma_{ijk}$, and the torsional SSC $t$.
  We classify magnetic textures into five categories depending on whether each geometric quantity vanishes (zero) or remains finite (\checkmark).
  “zero" means that a geometric quantity vanishes at all points in $\mathbb{R}^d$, and “\checkmark" means that it is finite at least at one point in $\mathbb{R}^d$.
  $d_\mathrm{p}$ is the corresponding parameter dimension for each class.
  }
  \label{table_classification}
  \begin{tabular}{|@{\hskip 5pt}c@{\hskip 5pt}||@{\hskip 5pt}c@{\hskip 5pt}|@{\hskip 5pt}c@{\hskip 5pt}|@{\hskip 5pt}c@{\hskip 5pt}|@{\hskip 5pt}c@{\hskip 5pt}|@{\hskip 5pt}c@{\hskip 5pt}|}
    \hline
    \multirow{2}{*}{} &
    \multirow{2}{*}{\makecell{Collinear magnets}} &
    \multirow{2}{*}{\makecell{Coplanar magnets}} &
    \multicolumn{3}{c|}{Noncoplanar magnets} \\
    \cline{4-6}
    & & & Type-I & Type-II & Type-III \\ \hline \hline
    \rotatebox{90}{~~~\makecell{Spin Textures}} &
    \includegraphics[width=0.15\textwidth]{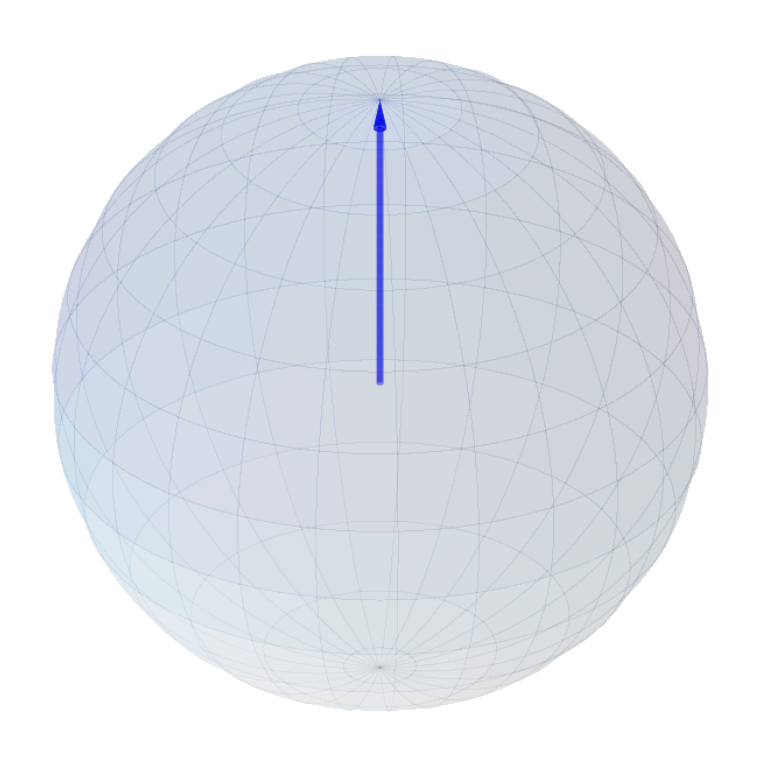} &
    \includegraphics[width=0.15\textwidth]{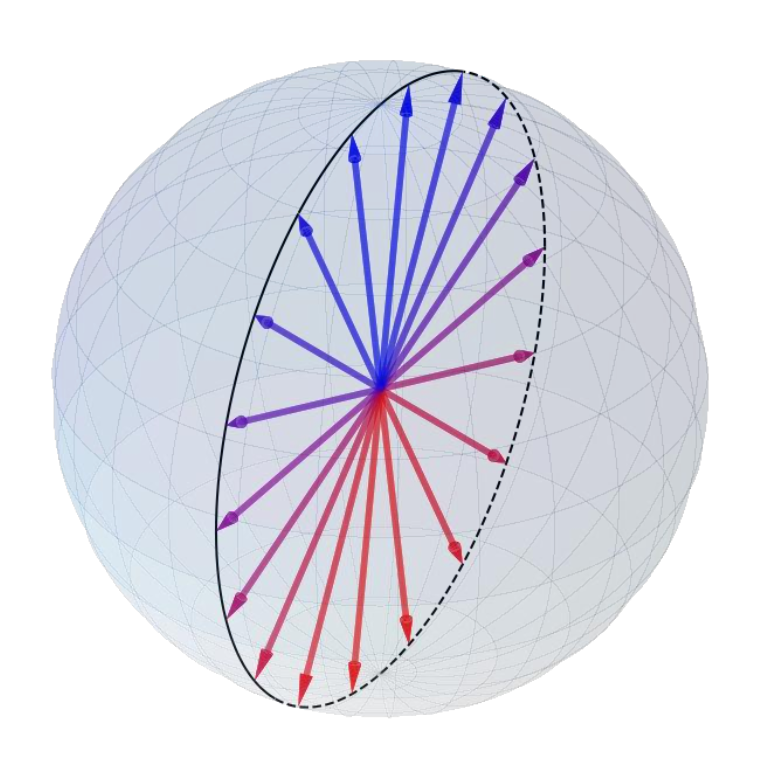} &
    \includegraphics[width=0.15\textwidth]{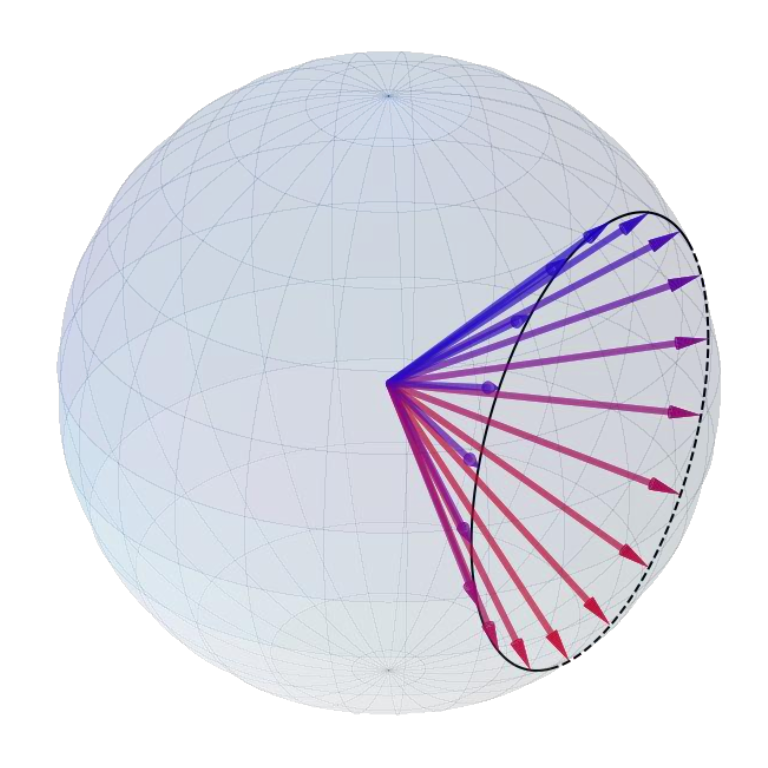} &
    \includegraphics[width=0.15\textwidth]{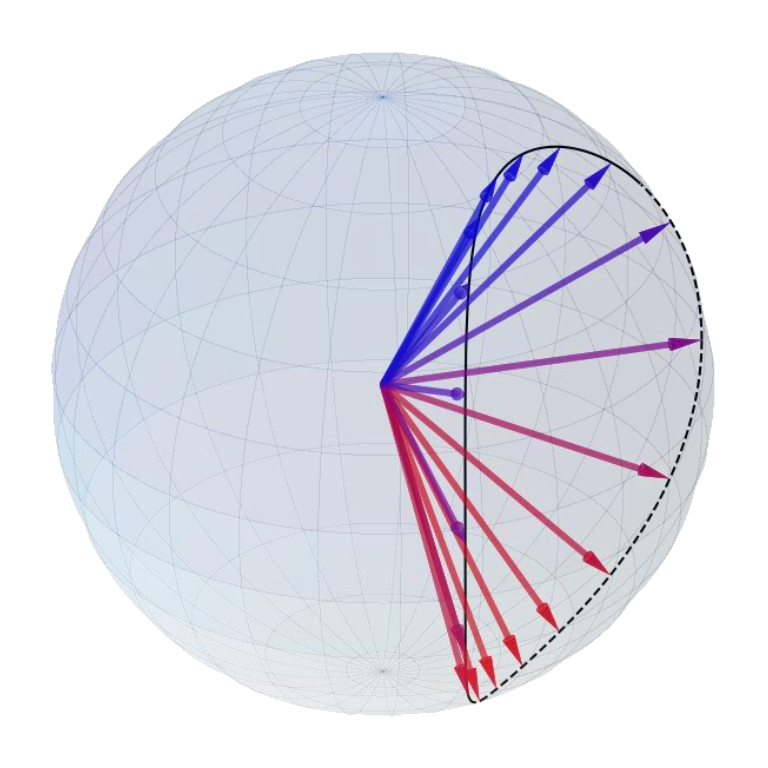} &
    \includegraphics[width=0.15\textwidth]{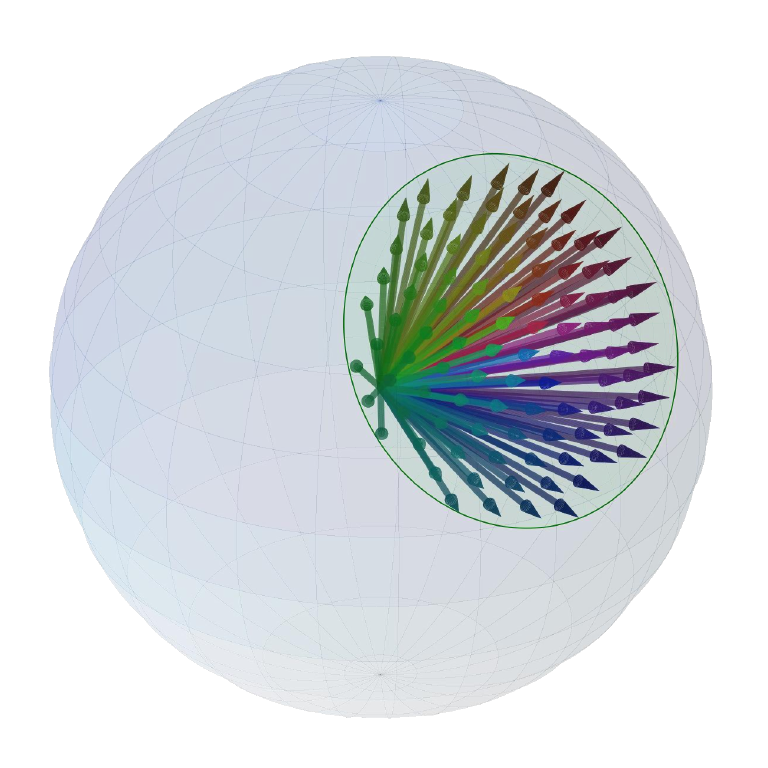} \\
    \hline
    $\bm{v}_i$ & zero & $\checkmark$ & $\checkmark$ & $\checkmark$ & $\checkmark$ \\
    \hline
    $\chi_{ij}$ & zero & zero & zero & zero & $\checkmark$ \\
    \hline
    $\gamma_{ijk}$ & zero & zero & $\checkmark$ & $\checkmark$ & $\checkmark$ \\
    \hline
    $t$ & zero & zero & zero & $\checkmark$ & -- \\
    \hline
    $d_\mathrm{p}$ & $d_\mathrm{p}=0$ & $d_\mathrm{p} = 1$ & $d_\mathrm{p} = 1$ & $d_\mathrm{p} = 1$ & $d_\mathrm{p} = 2$ \\ \hline
  \end{tabular}
\end{table*}

\subsection{Classification}
In the above discussion, we have introduced four differential geometric quantities, namely the vector spin chirality $\bm{v}_i$, the scalar spin chirality (SSC) $\chi_{xy}$, the geodesic SSC $\gamma_{ijk}$, and the torsional SSC $t$, that characterize the geometric nature of curves and surfaces traced by the spin $\bm{n}(\bm{x})$ on $\mathbb{S}^2$. 
Using these quantities, we propose a new framework for classifying magnetic textures, as shown in Table~\ref{table_classification}.
Broadly speaking, magnetic textures can be classified into collinear, coplanar, and noncoplanar magnets. In the following, we discuss each type of magnet in more detail.

\subsubsection{Collinear magnets}
Collinear magnets are characterized by the vanishing of all four geometric quantities.
When the VSC vanishes, the other three geometric quantities also vanish automatically.
A zero VSC implies that the spin does not trace any trajectory on $\mathbb{S}^2$ and remains fixed at a single point.
Therefore, the spin texture is given by
\begin{equation}
    \bm{n}^{\mathrm{cl}}(\bm{x}) = ( 0, 0, n_z ),
\end{equation}
which corresponds to a collinear magnetic texture.

\subsubsection{Coplanar magnets}
Coplanar magnets are characterized by a finite VSC and the vanishing of the other three geometric quantities.
When the SSC vanishes, the spin does not form a solid angle on $\mathbb{S}^2$, and its trajectory is restricted to a one-dimensional curve with $d_\mathrm{p} = 1$.
Furthermore, because the geodesic curvature $\kappa_g$ is zero, which automatically implies a vanishing torsion, the curve is confined to a great circle on $\mathbb{S}^2$.
Therefore, the spin texture is given by
\begin{equation}
    \bm{n}^{\mathrm{cp}}(\bm{x}) = ( \cos \Theta (\bm{x}) , \sin \Theta (\bm{x}) , 0 ),
\end{equation}
which depends on a single parameter $\Theta$ and represents a coplanar magnetic texture.
Coplanar magnets include helical magnets, fan magnets, cycloid magnets, chiral soliton lattices, and Bloch-type and N\'eel-type domain walls.

\subsubsection{Noncoplanar magnets: Type-I}
Since the SSC vanishes, the spin trajectory is restricted to a one-dimensional curve with $d_\mathrm{p} = 1$.
Moreover, because the geodesic curvature $\kappa_{g}$ is finite while the torsion $\tau$ is zero, the curve forms a small circle on $\mathbb{S}^2$.
Therefore, the spin texture is given by
\begin{equation}
    \bm{n}^{\mathrm{ncp}}_{\text{I}}(\bm{x})
    = ( \cos \alpha \cos \Theta(\bm{x}), \cos \alpha \sin \Theta (\bm{x}), \sin \alpha ),
\end{equation}
which depends on a single parameter $\Theta$.
Here, $\alpha$ satisfies $0 < \alpha < \pi$ and $\alpha \neq \pi/2$, and as long as this condition holds, the spins themselves do not lie in a single plane, resulting in a noncoplanar configuration.
In contrast, the curve itself is confined to a single plane.
Representative examples of this class include conical magnets.
In addition, this class includes noncoplanar spin textures obtained from coplanar magnets by applying an out-of-plane magnetic field.

\subsubsection{Noncoplanar magnets: Type-II}
Since the SSC vanishes, the spin trajectory is restricted to a one-dimensional curve with $d_\mathrm{p} = 1$.
When the torsion is finite, the curve is no longer planar and traces a generic curve on $\mathbb{S}^2$ without further restrictions.
The spin texture is given by
\begin{equation}
    \bm{n}^{\mathrm{ncp}}_{\text{II}}(\bm{x})
    =
    (n_x(\Theta(\bm{x})), n_y(\Theta(\bm{x})), n_z(\Theta(\bm{x}))),
\end{equation}
which depends on a single parameter $\Theta$ and satisfies the normalization condition $|\bm{n}^{\mathrm{ncp}}_{\text{II}}(\bm{x})| =1$ at each point.
An example is a distorted conical magnet as expressed in Eq.~(\ref{eq_dist_conical}).

So far, we have introduced three classes for the case $d_\mathrm{p}=1$.
It is well known that a curve in three-dimensional space is uniquely determined once the initial tangent vector, together with its curvature and torsion, is specified, which is known as the fundamental theorem of space curves.
When restricted to curves on $\mathbb{S}^2$, this uniqueness reduces to the specification of the initial condition and the geodesic curvature.
Therefore, in the classification of spherical curves with $d_\mathrm{p}=1$, the three classes introduced above provide a complete classification.

\subsubsection{Noncoplanar magnets: Type-III}
When a SSC is finite, the spin texture generically depends on two independent parameters and is expressed as
\begin{equation}
    \bm{n}^{\mathrm{ncp}}_{\text{III}}(\bm{x})
    =
    (n_x(\bm{\Theta}(\bm{x})), n_y(\bm{\Theta}(\bm{x})), n_z(\bm{\Theta}(\bm{x}))),
\end{equation}
which depends on just two parameters $\bm{\Theta} = (\Theta_1 , \Theta_2)$ and satisfies the normalization condition $|\bm{n}^{\mathrm{ncp}}_{\text{III}}(\bm{x})| = 1$ at each point.
This class includes single skyrmions and multi-Q magnetic states in skyrmion crystals, but it does not require the spin texture to be topologically nontrivial.

\vskip\baselineskip
The classification is based on whether, for each geometric quantity, there exists at least one point in $\mathbb{R}^d$ where it is finite or whether it vanishes at all points in $\mathbb{R}^d$.
As discussed above, when $\chi_{ij}$ is finite, $\gamma_{ijk}$ is also finite in many cases; however, their spatial averages are not necessarily finite.
Accordingly, it is possible to classify the type-III class into four classes depending on the presence or absence of $\braket{\chi_{ij}}$ and $\braket{\gamma_{ijk}}$, in the same manner as discussed in Table~\ref{table_ad_chi_gamma}.

There are several representative examples of materials belonging to the noncoplanar magnetic classes.
Noncentrosymmetric B20 compounds, such as MnSi, FeGe, and $\mathrm{Fe}_{1-x}\mathrm{Co}_x\mathrm{Si}$, exhibit multiple magnetic phases in the magnetic-field–temperature phase diagram \cite{muhlbauer2009skyrmion}. 
The low-field region hosts a helical phase, while the high-field and low-temperature region hosts a conical phase (Type-I). 
Near the transition temperature, a skyrmion-lattice phase also appears (Type-III).
A similar phase diagram has also been reported in centrosymmetric crystals such as $\mathrm{Gd}_3\mathrm{Ru}_4\mathrm{Al}_12$, where conical phases (Type-I) and skyrmion-lattice phases (Type-III) have been observed \cite{hirschberger2019skyrmion}.
Even when a skyrmion-lattice phase has not been identified, centrosymmetric crystals exhibiting a Type-I phase have been reported, including MnP \cite{PhysRevB.77.104439}, $\alpha$-$\mathrm{EuP}_3$ \cite{doi:10.1073/pnas.2405839122}, and $\mathrm{YMn}_6\mathrm{Sn}_6$ \cite{https://doi.org/10.1002/adma.202420614}.
In addition, distorted conical phases (Type-II) have been reported in several systems, including $\mathrm{Cu}_2\mathrm{OSeO}_3$ \cite{mehboodi2025observation} and MnSi thin films \cite{PhysRevB.85.094429}.

\section{Semiclassical theory of magnetic textures} \label{sec_semiclassical}
We have classified magnetic textures using geometric quantities that play central roles in differential geometry.
Among these quantities, the SSC acts as an effective magnetic field within the semiclassical theory \cite{PhysRevLett.83.3737,PhysRevLett.93.096806,PhysRevLett.69.3593,RevModPhys.82.1959,PhysRevB.59.14915} and influences electron dynamics through a Lorentz force.
As a consequence, electrons moving on a skyrmion exhibit the topological Hall effect, with the Hall conductivity proportional to the skyrmion number, which is given by the spatial integral of the SSC.
Remarkably, as discussed below, other geometric quantities such as the Riemannian metric and the geodesic SSC also appear in the semiclassical dynamics of electrons.
Conventional semiclassical theory is based on the adiabatic approximation and assumes that the spatial modulation of the spin texture is very slow, retaining only the lowest-order gradient terms.
These new geometric quantities arise from nonadiabatic effects and from contributions of higher-order gradients.

\subsection{Quantum effective low-energy Hamiltonian}
We consider a free-electron gas model coupled to classical localized spins $\bm{n}(\bm{x})$ with spatial modulation. We assume that the magnitude of the spins is unity ($|\bm{n}(\bm{x})| = 1$), and we do not consider spatial modulations of the amplitude, such as those in spin density wave states.
The Hamiltonian is given by
\begin{equation} \label{eq_hamiltonian}
    \hat{H} = \frac{(\hat{\bm{p}} + e\bm{A}(t) )^2}{2m} - J \bm{n} (\hat{\bm{x}}) \cdot \bm{\sigma}.
\end{equation}
Here, $-e(<0)$ is the electron charge, and $m$ is the electron mass. 
$\hat{\bm{p}}$ is the momentum operator, $\hat{\bm{x}}$ is the position operator satisfying the canonical commutation relations $[\hat{x}_i, \hat{p}_j] = i\hbar \delta_{ij}$, and $\bm{\sigma}$ are the Pauli matrices describing the spin degrees of freedom of electrons.
$\bm{A}(t)$ denotes the vector potential describing the electric field. $J$ is the exchange coupling constant between the localized spins and the conduction electrons.

After performing a unitary transformation to the local spin frame, the Hamiltonian becomes
\begin{equation}
    \hat{H}' = U^\dagger(\hat{\bm{x}}) \hat{H} U(\hat{\bm{x}}) = \frac{1}{2m} (\hat{\bm{p}} + e \bm{A}(t) - \bm{a}(\hat{\bm{x}}))^2 - J \sigma_z,
\end{equation}
where $\bm{a} = i \hbar U^\dagger \bm{\nabla} U$ is a pure U(2) gauge field.
The trace part of this gauge field can be absorbed into the vector potential. Since the 
magnetic field is not considered in this work, the trace part is neglected in the following. Consequently, $\bm{a}$ is a pure $\mathrm{SU}(2)$ gauge field.

In the following, we assume that $J$ is the largest energy scale and treat the kinetic-energy term perturbatively within the low-energy ($-J$) sector. 
The $\mathrm{SU}(2)$ gauge field generally contains off-diagonal components, which induce mixing with the high-energy ($+J$) sector and give rise to nonadiabatic effects.
This perturbation is valid when $(E_F/J) \cdot (\lambda_F / \lambda_s) \ll 1$, where $E_F$ is the Fermi energy, $\lambda_F$ is the Fermi wavelength, and $\lambda_s$ is the characteristic length scale of the spatial modulation of the spin texture.
Within the approximation up to order $J^{-2}$, the effective Hamiltonian is given by (see details in Appendix \ref{appendix_quantum_ham})
\begin{equation}
    \hat{H}_{\mathrm{eff}} = - J + \hat{H}_{\mathrm{ad}} + \hat{H}_{\mathrm{nad},1} +  \hat{H}_{\mathrm{nad},2} + \mathcal{O}(J^{-3}).
\end{equation}
The second term is the effective Hamiltonian within the adiabatic approximation and is expressed as \cite{PhysRevLett.93.096806,PhysRevLett.69.3593}
\begin{equation}
    \hat{H}_{\mathrm{ad}} = \frac{1}{2m} ( \hat{\bm{\pi}}_1^2 + |\bm{a}_{12}|^2 ).
\end{equation}
The third term is the first-order nonadiabatic correction and is given by
\begin{equation}
    \hat{H}_{\mathrm{nad},1} = - \frac{ \hat{\Lambda} \hat{\Lambda}^\dagger }{2J},
\end{equation}
which is of order $J^{-1}$ \cite{qxnw-8q4y}.
The fourth term, $\hat{H}_{\mathrm{nad},2}$, is a new second-order nonadiabatic correction, and the expression is
\begin{equation}
    \hat{H}_{\mathrm{nad},2} = \frac{1}{8J^2} ( \{ \hat{H}_1, \hat{\Lambda} \hat{\Lambda}^\dagger \} - 2 \hat{\Lambda} \hat{H}_2 \hat{\Lambda}^\dagger ),
\end{equation}
which is of order $J^{-2}$.
Here, $\hat{\bm{\pi}} = \hat{\bm{p}} + e\bm{A}(t)$, $\hat{\bm{\pi}}_{\alpha(=1,2)} = \hat{\bm{\pi}} - \bm{a}_{\alpha \alpha} = \hat{\bm{\pi}} \mp \bm{a}_z$, $\bm{a}_z = (\bm{a}_{11} - \bm{a}_{22})/2$, $\hat{\Lambda} = -\frac{1}{2m} \{ \hat{\pi}_i , a_{12i} \}$, and $\hat{H}_\alpha = \frac{1}{2m} ( \hat{\bm{\pi}}_\alpha^2 + |\bm{a}_{12}|^2 )$.
$\{ A , B\} =AB + BA$ is the anticommutator.

\subsection{Semiclassical Hamiltonian}
We now derive the semiclassical Hamiltonian corresponding to the quantum effective low-energy Hamiltonian obtained above. 
To map operators onto their classical counterparts in phase space, we employ the Wigner transformation \cite{PhysRev.40.749,zachos2005quantum}, which is defined as
\begin{equation}
    A_W(\bm{x} , \bm{p}) = \int d^dy e^{ -i \bm{p} \cdot \bm{y}} \braket{ \bm{x}+\bm{y}/2 | \hat{A} | \bm{x} - \bm{y}/2  }.
\end{equation}
Here, $(\bm{x},\bm{p})$ are the canonical variables (phase-space coordinates).
One of the important properties of the Wigner transformation is that the Wigner transformation of a product of operators corresponds to the Moyal product \cite{moyal1949quantum,zachos2005quantum},
\begin{equation}
    \hat{A} \hat{B} \to A_W \star B_W,~~~\star = \exp \Bigl(\frac{i \hbar}{2} ( \overleftarrow{\partial_{\bm{x}}} \cdot \overrightarrow{\partial_{\bm{p}}} - \overleftarrow{\partial_{\bm{p}}} \cdot \overrightarrow{\partial_{\bm{x}}}  )  \Bigr).
\end{equation}
The $\hbar$ expansion of the Moyal product corresponds to a gradient expansion.
For the product of identical operators whose dependence on $\bm{x}$ or $\bm{p}$ is at most linear, the gradient expansion of the Moyal product does not generate terms beyond the first order in $\hbar$ and thus reduces to the ordinary product. 
Consequently, the Hamiltonian in the adiabatic approximation is obtained simply by replacing operators with their classical counterparts.
By contrast, nonadiabatic effects give rise to contributions beyond the first order in the $\hbar$ expansion. 
These contributions give rise to a variety of quantum geometric quantities that go beyond the real-space Berry curvature discussed in the conventional adiabatic semiclassical theory, as explicitly shown below.

After performing the $\hbar$ expansion, we obtain the semiclassical Hamiltonian.
The semiclassical Hamiltonian for the adiabatic case is (see detailed derivations in Appendix \ref{appendix_semi_ham})
\begin{equation} \label{eq_semihamiltonian_ad}
    H^{\mathrm{cl}}_{\mathrm{ad}} = \frac{1}{2m} ( \bm{\pi}_1^2 + \mathrm{tr} G ),
\end{equation}
which is a straightforward replacement of operators by their classical counterparts.
The semiclassical Hamiltonian for the first-order nonadiabatic correction is 
\begin{equation} \label{eq_semihamiltonian_ad_1}
    H^{\mathrm{cl}}_{\mathrm{nad},1} = -\frac{1}{2m^2 J} ( G_{ij} \pi_{1i} \pi_{1j} + \Gamma_i \pi_{1i} + R ),
\end{equation}
where the first term is obtained by the simple replacement from operators to classical counterparts \cite{qxnw-8q4y}.
The second and third terms are new terms obtained from the $\hbar$ expansion.
The semiclassical Hamiltonian for the second-order nonadiabatic correction up to order $\lambda^{-3}_s$ is
\begin{equation} \label{eq_semihamiltonian_ad_2}
    H^{\mathrm{cl}}_{\mathrm{nad},2} = -\frac{1}{4m^3 J^2} \Gamma_{ijk} \pi_{1i} \pi_{1j} \pi_{1k}.
\end{equation}
Here, the zeroth-order term in the $\hbar$ expansion cancels out, and the expansion starts at order $\hbar^1$.
Unlike the adiabatic Hamiltonian in Eq.~(\ref{eq_semihamiltonian_ad}), these nonadiabatic corrections in Eqs.~(\ref{eq_semihamiltonian_ad_1}) and (\ref{eq_semihamiltonian_ad_2}) include terms beyond the simple replacement arising from higher-order gradients generated by the $\hbar$ expansion, accompanied by quantum geometric quantities that extend beyond $G$.
Here, various quantum geometric quantities emerge, including 
\begin{equation}
    \begin{aligned}
        &G_{ij} = \mathrm{Re}[ a_{12i} a^*_{12j} ], \\
        &\Gamma_{ijk} = \mathrm{Im}[ a_{12i} D_k a^*_{12j} ], \\
        &R = \frac{1}{4} \mathrm{Re}[ D^*_j a_{12i} D_i a_{12j}^*  ].
    \end{aligned}
\end{equation} 
$D_i = \hbar \partial_i + 2ia_{zi}$ is the covariant derivative.
In addition, $\Gamma_i = \Gamma_{jij}$.
Because $G$ is semipositive definite, the $\mathrm{tr} G$ term gives a repulsive potential, and the $G_{ij} \pi_{1i} \pi_{1j}$ term increases the effective mass.
This $G$ term produces an emergent gravity described by the geodesic equation \cite{qxnw-8q4y,pwvz-868b,yoshida2025,maranzana2026}.
Furthermore, the $\Gamma$ terms indicate that the spin texture effectively induces an asymmetry in the band structure and also provides an asymmetric effective mass tensor between $\bm{\pi}$ and $-\bm{\pi}$.

It can be explicitly verified that these quantum geometric quantities are gauge invariant. 
In particular, the term $-J\sigma_z$ in Eq.~(\ref{eq_hamiltonian}) remains invariant under both a $\mathrm{U}(1)$ gauge transformation $e^{i\theta(\bm{x})}$ and a $\mathrm{U}_z(1)$ gauge transformation $e^{i\chi(\bm{x}) \sigma_z}$.
Under each of these gauge transformations, the $\mathrm{SU}(2)$ gauge field transforms as
\begin{equation}
    \begin{aligned}
        &\bm{a} \overset{\mathrm{U}(1)}{\to} \bm{a} - \hbar\bm{\nabla} \theta, \\
        &\bm{a} \overset{\mathrm{U}_z(1)}{\to}
        \begin{pmatrix}
            \bm{a}_{11} - \hbar\bm{\nabla} \chi & \bm{a}_{12} e^{-2i\chi} \\
            \bm{a}^*_{12} e^{+2i\chi} & \bm{a}_{22} + \hbar\bm{\nabla} \chi
        \end{pmatrix}.
    \end{aligned}
\end{equation}
Under a $\mathrm{U}(1)$ gauge transformation, the off-diagonal and $z$-components of the $\mathrm{SU}(2)$ gauge field remain unchanged, and thus the gauge invariance is manifest. 
By contrast, for a $\mathrm{U}_z(1)$ gauge transformation, the properties of the covariant derivative $D_i$ play a crucial role.
For example, the transformation of $\Gamma_{ijk}$ is given by
\begin{align}
    \Gamma_{ijk} &\overset{\mathrm{U}_z(1)}{\to} \mathrm{Im}[ a_{12i} e^{-2i \chi} ( \hbar\partial_k + 2i a_{zk} - 2i \partial_k \chi  ) e^{2i\chi} a^{*}_{12j} ] \nonumber \\
    &~~~~~~~=\Gamma_{ijk}.
\end{align}
Thus, it is gauge independent.
Similarly, gauge invariance can be checked for the other quantum geometric quantities as well.

These quantum geometric quantities can be expressed explicitly in terms of the spin texture $\bm{n}(\bm{x})$, and are found to coincide with the geometric quantities discussed in Sec.~\ref{sec_differential_geometry} up to overall prefactors (see detailed derivations in Appendix \ref{appendix_spin_rep}):
\begin{equation}
\label{eq_G_Gamma_spin}
    \begin{aligned}
        &G_{ij} = \frac{\hbar^2}{4} \partial_i \bm{n} \cdot \partial_j \bm{n} = \frac{\hbar^2}{4} g_{ij}, \\
        &\Gamma_{ijk} = \frac{\hbar^3}{4} \bm{n} \cdot ( \partial_i \bm{n} \times \partial_{jk} \bm{n} ) = \frac{\hbar^3}{4} \gamma_{ijk}. 
    \end{aligned}
\end{equation}

\subsection{Asymmetric band structure by $\Gamma_{ijk}$}
We evaluate the band asymmetry induced by the geodesic SSC $\Gamma_{ijk}$.
Here, we consider a spatially one-dimensional conical magnet in $\mathbb{R}^{d=1}$ described by Eq.~(\ref{eq_conical}), and the minimal Hamiltonian inducing the band asymmetry with the asymmetric effective mass is given by
\begin{equation}
    H(x,\pi) = \frac{\pi^2}{2m} -\frac{\Gamma_{xxx}}{4m^3J^2}  \pi^3.
\end{equation}
Although the $\Gamma$-induced band asymmetry already appears at order $J^{-1}$ in Eq.~(\ref{eq_semihamiltonian_ad_1}), it does not generate an asymmetry in the effective mass tensor and is therefore neglected. Moreover, as will be clarified in the following discussion, the $J^{-1}$ term can be gauged out and does not appear in the electron dynamics, and is thus not essential.
The band asymmetry is defined as
\begin{equation}
    \Delta_\mathrm{BA}(\pi) = \frac{|\braket{H}(\pi) - \braket{H}(-\pi)|}{\braket{H}(\pi) + \braket{H}(-\pi)},
\end{equation}
where we use the notation $\braket{O} = \int d^d\bm{x} O/V$ for the spatial average.
The band asymmetry for the spatially one-dimensional conical magnet at the Fermi surface ($\pi = \pi_F = \hbar k_F$, where $k_F$ is the Fermi wavenumber) reads
\begin{equation}
    \Delta_{\mathrm{BA}}(\pi_F) = \frac{\hbar^4 k_F |q^3 \cos^2 \alpha \sin \alpha| }{8m^2J^2}.
\end{equation}
For the bare mass $m = 9.1 \times 10^{-31}~\mathrm{kg}$, the coupling constant $J = 0.1~\mathrm{eV}$, the Fermi wavenumber $k_F = 1~\mathrm{\AA^{-1}}$, and the wavenumber of the spin modulation $q = q_0~\mathrm{nm^{-1}}$, we obtain $\Delta_{\mathrm{BA}} = 7.0\cdot q_0^3 |\cos^2 \alpha \sin \alpha|$.
In the case of $q = 0.1~\mathrm{nm^{-1}}$, the band asymmetry reaches at most $0.028 \%$.
Note that, since the band asymmetry scales as $q^3$, a one-order-of-magnitude change in $q$ leads to a three-order-of-magnitude change in the band asymmetry.
\begin{figure}[t]
\centering
\includegraphics[width=0.9 \linewidth]{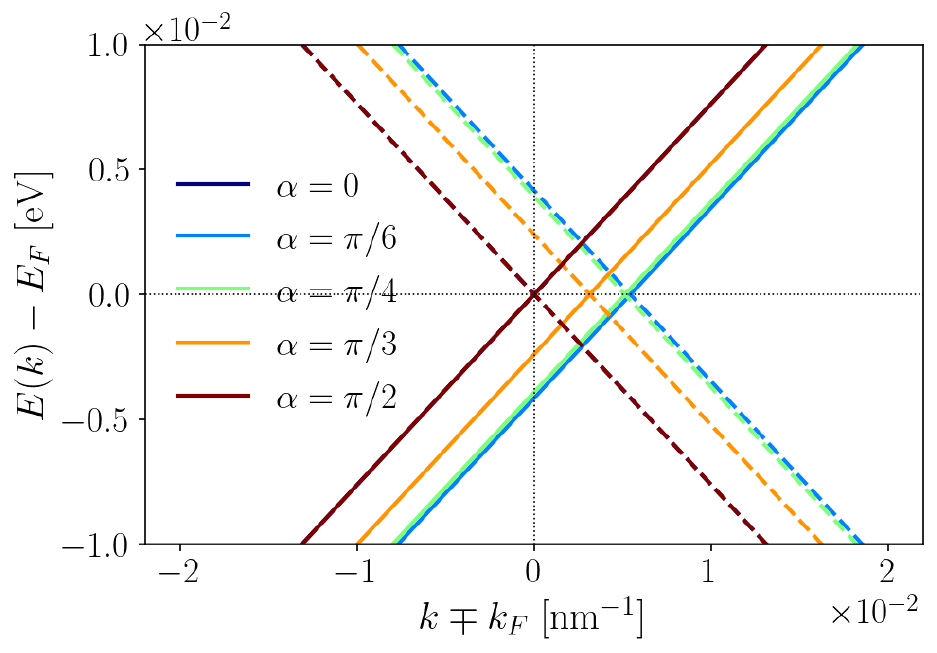}
\caption{
Energy dispersion $E(k) = \braket{H}(\hbar k)$ for the conical magnet. We fix $k_F = 1 \mathrm{\AA}^{-1}$ and $E_F = \hbar^2 k_F^2 / 2m$, and the energy dispersion is shown near $\pm k_F$ and $E_F$.
The solid and dashed lines correspond to the energy dispersions near $+k_F$ and $-k_F$, respectively.
} \label{fig_asymmetry}
\end{figure}
We display the energy dispersion in the vicinity of the Fermi surface in Fig.~\ref{fig_asymmetry}.
For $0 < \alpha < \pi/2$, the energy dispersion becomes asymmetric between $k$ and $-k$ (here, $\pi = \hbar k$).
On the other hand, for $\alpha =0$ (helical magnets) and $\pi/2$ (collinear ferromagnets), the band dispersions are symmetric because $\Gamma_{xxx}$ vanishes.
The band asymmetry in conical magnets has been discussed in Ref.~\cite{lhj4-9h29,doi:10.1073/pnas.2405839122,doi:10.1126/sciadv.adw8023,https://doi.org/10.1002/adma.202420614}.
Here, however, we emphasize that it is sufficient for $\Gamma_{ijk}$ to be finite, and that the band asymmetry can arise in more general situations beyond conical magnetic textures, including
the skyrmionic textures discussed in Sec.~\ref{sec_two_dim} and other noncoplanar magnets discussed in Refs.~\cite{doi:10.7566/JPSJ.91.094704,PhysRevB.106.014420,solids6030040}.

\section{Emergent Responses related to geometry}
In this section, we discuss emergent responses arising from the Riemannian metric $G_{ij}$ and the geodesic scalar spin chirality (geodesic SSC) $\Gamma_{ijk}$ induced by nonadiabatic effects.

\subsection{Resistivity change induced by the Riemannian metric}
To focus on the effects of the Riemannian metric, we retain terms up to second order in the spatial gradients and first order in nonadiabatic effects.
The resulting semiclassical Hamiltonian is (in the following, we denote $\bm{\pi}_1$ simply as $\bm{\pi}$.)
\begin{equation}
    H = \frac{\bm{\pi}^2}{2m} - \bar{G}_{ij} \pi_i \pi_j + V,
\end{equation}
where $\bar{G}_{ij} =  G_{ij} / 2m^2 J$ and $V = \mathrm{tr} G / 2m$.
The canonical equations read
\begin{equation} \label{eq_canonical_1}
    \begin{aligned}
        \dot{x}_i &= \frac{\pi_i}{m} -2 \bar{G}_{ij} \pi_j, \\
        \dot{\pi}_i &=
        F_{ij} \dot{x}_j -eE_i + \mathcal{O}(\lambda^{-3}_s).
    \end{aligned}
\end{equation}
Here, $F_{ij} (= \partial_i a_{zj} - \partial_j a_{zi})$ is the real-space Berry curvature, which gives rise to the Lorentz force.
By solving the Boltzmann equation with the relaxation time approximation
\begin{equation}
    \dot{\pi}_i \partial_{\pi_i} f + \dot{x}_i \partial_{x_i} f = - \frac{f-f^{\mathrm{eq}}}{\tau},
\end{equation}
we evaluate the current under the electric field $\bm{E}$, using 
\begin{equation}
    j_i = -e \int_{\bm{x},\bm{\pi}}  \dot{x}_i f.
\end{equation}
Here, we define $\int_{\bm{x},\bm{\pi}} \equiv  \int d^d\bm{x} /V \int d^d\bm{\pi}/(2\pi)^d $.
$f^{\mathrm{eq}}(H) = 1/( 1 + e^{\beta (H - \mu )} )$ is the local thermal equilibrium distribution function with an inverse temperature $\beta$ and a chemical potential $\mu$, and $f$ is a nonequilibrium distribution function. $\tau$ is the relaxation time.
We solve this equation to linear order in the electric field and to second order in spatial gradients, assuming that the magnetic texture varies slowly in space ($\lambda_F / \lambda_s \ll 1$).
Using the resulting distribution function to evaluate the current, we find that the electrical conductivity $\sigma_{ij}$ is given by three terms (see detailed derivations in Appendix \ref{appendix_linear_reimann})
\begin{equation} \label{eq_conductivity}
    \sigma_{ij} = \sigma_{ij}^{\mathrm{L},0} + \sigma_{ij}^{\mathrm{L},G} + \sigma_{ij}^{\mathrm{H}}.
\end{equation}
The first term $\sigma_{ij}^{\mathrm{L},0} = \tau e^2 n \delta_{ij} / m$ is the conventional longitudinal electrical conductivity, with the electron density $n = \int_{\bm{\pi}} f_0$ with $f_0 = f_{\mathrm{eq}}(\bm{\pi}^2/2m)$.
The second term $\sigma_{ij}^{\mathrm{L},G}$ is the correction of the longitudinal electrical conductivity arising from the Riemannian metric.
This correction is given by
\begin{equation} \label{eq_cond_LG}
    \sigma_{ij}^{\mathrm{L},G} = - \frac{ \tau e^2 n }{m^2 J} \Bigl( \braket{G_{ij}} -\frac{\delta_{ij} }{2} \mathrm{tr} \braket{G} \Bigr) - \frac{\tau e^2 n' \mathrm{tr} \braket{G}}{2m^2} \delta_{ij}.
\end{equation}
Here, $n' = \partial n / \partial \mu$.
The first term in Eq.~(\ref{eq_cond_LG}) arises from nonadiabatic effects at order $J^{-1}$, whose mechanism corresponds to the mass enhancement,
whereas the second term is independent of $J$ and is obtained within the adiabatic approximation, whose mechanism corresponds to the repulsive potential.
The third term
\begin{equation} \label{eq_cond_H}
\sigma^\mathrm{H}_{ij} = \frac{\tau^2 e^2 n}{m^2} \braket{F_{ij}} 
\end{equation}
is the transverse electrical conductivity known as the topological Hall effect (THE).
This Hall conductivity is independent of $J$ and is obtained within the adiabatic approximation.

\begin{figure}[t]
\centering
\includegraphics[width=1.0 \linewidth]{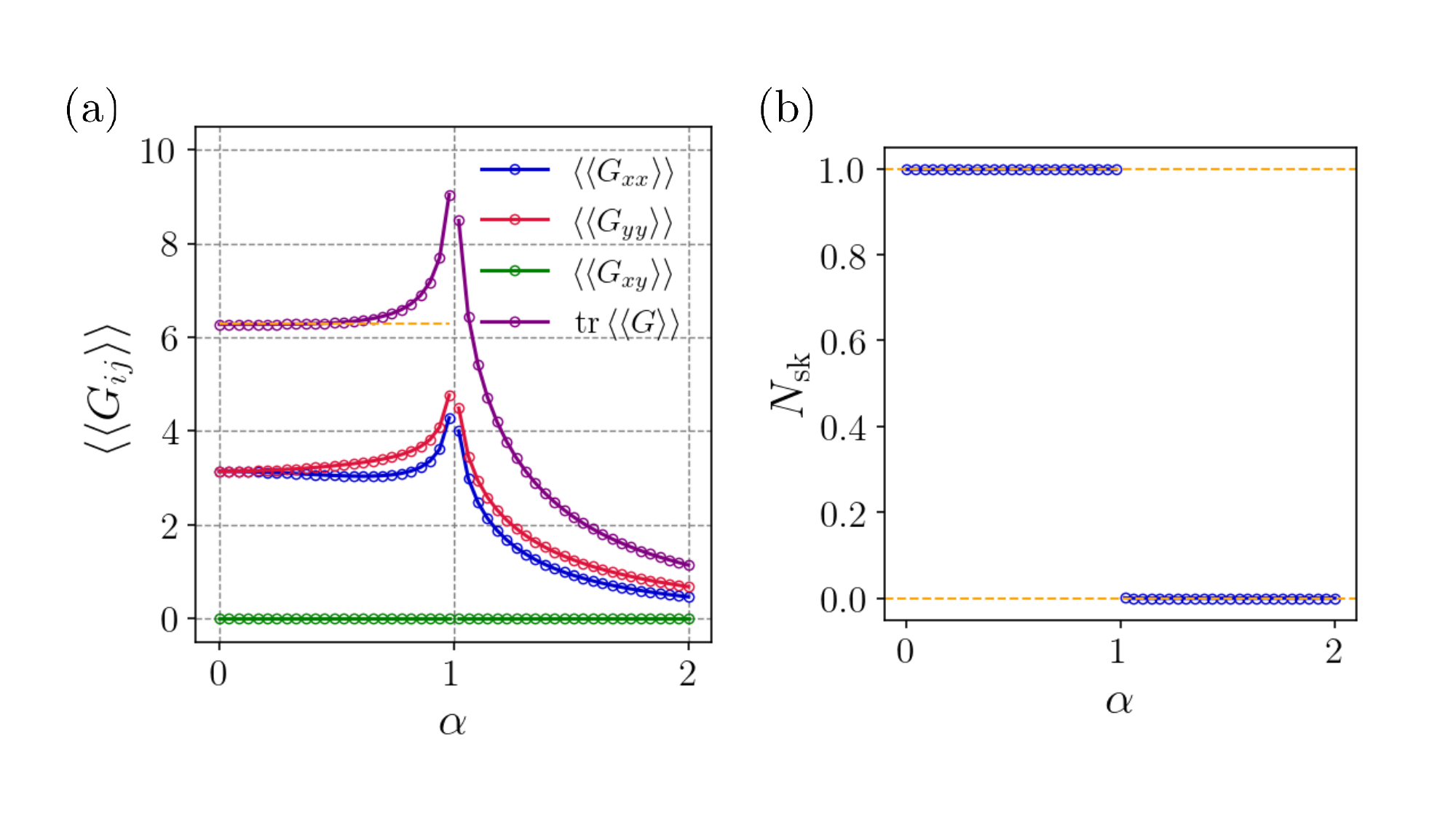}
\caption{
(a) The $\alpha$ dependence of the spatial integral of the Riemannian metric $\braket{\braket{G_{ij}}}$ in a two-dimensional skyrmion. The orange dashed line corresponds to $2\pi N_{\mathrm{sk}}$.
Here, we set $\hbar=1$.
(b) The $\alpha$ dependence of the skyrmion number.
The topological transition occurs at $\alpha =1$.
} \label{fig_metric}
\end{figure}

As can be seen from its spin representation in Eq.~(\ref{eq_G_Gamma_spin}), the Riemannian metric corresponds to the exchange energy of spins and is expected to be enhanced when the spatial modulation of spin textures is steep, resulting in a large change in the electrical resistivity.
To verify this point, we consider a two-dimensional skyrmion and introduce a model in which a transverse magnetic field is applied to forcibly drive a topological transition.
We consider a two-dimensional skyrmion in the absence of a magnetic field, with the form
$\bm{n}(r,\phi) = ( \sin \Theta (r) \cos \Phi(\phi) , \sin \Theta (r) \sin \Phi(\phi) , \cos \Theta(r) )$
with polar coordinates $(r, \phi)$.
Here, we define $\Theta(r) = 2\arctan r^{-1}$ and $\Phi(\phi) = m \phi + \gamma$ with a vorticity $m$ and a helicity $\gamma$.
This skyrmionic texture is obtained by minimizing the exchange energy $E_{\mathrm{ex}} =  \int  d^2\bm{x} \partial_i \bm{n} \cdot \partial_i \bm{n}$ \cite{belavin1975metastable}.
Here, we consider that a magnetic field is applied along the $x$ direction, and assume that the spin is transformed as $\bm{n} \to \bm{N}/|\bm{N}|~(\bm{N} = \bm{n} + \alpha \bm{e}_x)$.
We treat $\alpha$ as an effective magnetic field.
In the following, we use the parameters $m = -1$ and $\gamma=0$.
In this model, a topological transition occurs at $\alpha=1$, yielding a nontrivial phase in the low-field regime ($N_{\mathrm{sk}} = 1$) and a trivial phase in the high-field regime ($N_{\mathrm{sk}} = 0$), as shown in Fig.~\ref{fig_metric}(b).
Figure \ref{fig_metric}(a) shows the $\alpha$ dependence of the spatial integral of the Riemannian metric $\braket{\braket{G_{ij}}} = \int d^2\bm{x} G_{ij}$.
In two-dimensional cases, the spatial integral of the Riemannian metric is bounded by the skyrmion number \cite{belavin1975metastable}, as
\begin{equation}
    \mathrm{tr}\braket{\braket{G}} \geq 2\pi \hbar^2 N_{\mathrm{sk}}.
\end{equation}
The orange dashed line in Fig.~\ref{fig_metric}(a) corresponds to the skyrmion number, and it clearly demonstrates the bound.
In particular, in the absence of a magnetic field, the equality between the two quantities is exactly satisfied.
For a topologically nontrivial texture, the spin configuration necessarily involves spatial modulation, 
so it is intuitive that the exchange energy and the spatial integral of the Riemannian metric take finite values.
The claim here is stronger: these quantities must always be equal to or greater than the skyrmion number.

Near the topological phase transition point, the spatial integral of the Riemannian metric is strongly enhanced and exhibits divergent behavior.
Near the transition point, most spin components tend to align with the direction of the magnetic-field, 
whereas components that are nearly antiparallel to the field do not readily rotate toward the field direction.
As a result, the spin configuration develops a steep spatial modulation at the interface between these regions,
which significantly enhances the spatial variation.
This tendency is expected to be generic to topological phase transitions.
For example, a similar enhancement should occur in chiral soliton systems realized by applying an in-plane magnetic field to helical magnets.

Using Eq.~(\ref{eq_cond_LG}), we estimate the change in the electrical conductivity induced by the spatial modulation of the spin texture.
We estimate the contributions from the first and second terms in Eq.~(\ref{eq_cond_LG}) separately.
For a two-dimensional system, the ratio between the first term of $\sigma^{\mathrm{L},G}_{xx}$ and the conventional contribution $\sigma^{\mathrm{L},0}_{xx}$ is $\Delta^{(1)} = \braket{\braket{G_{xx}-G_{yy}}} n_{\mathrm{sk}} / 2mJ$.
In the case of $J = 0.1~\mathrm{eV}$, $\hbar^2 / 2m J = 0.38~\mathrm{nm^2}$.
$n_{\mathrm{sk}}$ is the skyrmion density, and it typically takes values ranging from $10^{-5}~\mathrm{nm^{-2}}$ to $10^{-3}~\mathrm{nm^{-2}}$.
Thus, by using $\braket{\braket{G_{ii}}} \approx \mathcal{O}(1) \cdot \hbar^2$, the change $\Delta^{(1)}$ is expected to be on the order of $10^{-3}$-$10^{-1}\%$.
On the other hand, the ratio between the second term and the conventional contribution is $\Delta^{(2)} = n' \mathrm{tr} \braket{\braket{G}} n_{\mathrm{sk}} / 2nm$.
The coefficient $\hbar^2 n' /2nm \sim \hbar^2 / 2 m E_F$ is approximated to be $0.038~\mathrm{nm^2}$, assuming $E_F =1~\mathrm{eV}$.
Compared with $\Delta^{(1)}$, the energy scale in $\Delta^{(2)}$ is changed from $J$ to $E_F$.
Thus, the ratio $\Delta^{(2)}$ is reduced by approximately one order of magnitude.

\begin{table*}[t]
\centering
\caption{Summary table of emergent responses induced by geometric quantities such as the Riemannian metric, the scalar spin chirality (real-space Berry curvature), and the geodesic SSC.
The third row indicates the magnetic structures that can render each emergent response finite.
The fourth row summarizes whether each emergent response originates from adiabatic or nonadiabatic effects, and expresses the corresponding perturbative order in powers of $J$.
The fifth row indicates the order of gradients, and the seventh row shows whether each response is allowed (\checkmark) or forbidden (--) in the presence of spatial inversion symmetry.
}  \label{table_responses} 
\renewcommand{\arraystretch}{1.3}
\begin{tabular}{@{\hskip 5pt}c@{\hskip 5pt}||@{\hskip 5pt}c@{\hskip 5pt}@{\hskip 5pt}c@{\hskip 5pt}@{\hskip 5pt}c@{\hskip 5pt}} 
    \hline 
    Geometry / chirality & Riemannian metric $G$ & SSC (Berry curvature) $F$ & Geodesic SSC $\Gamma$  \\ \hline
    Emergent Response & Resistivity change & Topological Hall effect & Nonreciprocal response \\
    Spin texture & Noncollinear & Noncoplanar (Type-III) & Noncoplanar (All types) \\
    Adiabatic/Nonadiabatic & Both effects ($J^{0,-1}$) & Adiabatic effect ($J^0$) & Nonadiabatic effect ($J^{-2})$ \\
    Order of gradients &  $\lambda_s^{-2}$ & $\lambda_s^{-2}$ & $\lambda_s^{-3}$ \\
    Mechanism & \makecell{Repulsive potential/ \\ Mass enhancement}  & Effective orbital magnetic field & Effective band asymmetry \\
    $\mathcal{P}$-symmetry & \checkmark & \checkmark & -- \\
    Equation & Eq.~(\ref{eq_cond_LG}) & Eq.~(\ref{eq_cond_H}) & Eq.~(\ref{eq_cond_nr}) \\ \hline
\end{tabular}
\end{table*}

\subsection{Emergent nonreciprocal transport induced by the geodesic SSC}
As discussed above, the geodesic SSC renders the band dispersion asymmetric. 
Here, we analyze a nonreciprocal response arising from this asymmetry. 
At first order in nonadiabatic effects, the contribution from $\Gamma$ can be gauged away by a change of variables in $\pi$ and therefore does not appear in the response. 
In contrast, at second order in nonadiabatic effects, a cubic term in the dispersion emerges, which cannot be gauged out and thus contributes to the response coefficients. 
As the simplest model that incorporates such asymmetry, we consider
\begin{equation} \label{eq_Hamiltonian_Gamma_asymmetry}
    H = \frac{\bm{\pi}^2}{2m} - \bar{\Gamma}_{(ijk)} \pi_i \pi_j \pi_k,
\end{equation}
where we renormalize the coefficient such that $\bar{\Gamma}_{ijk} = \Gamma_{ijk}/ 4m^3 J^2$.
The parentheses $(ijk)$ indicate full symmetrization over the indices.

The canonical equations read
\begin{equation}
    \begin{aligned}
        \dot{x}_i &= \frac{\pi_i}{m} - 3 \bar{\Gamma}_{(ijk)} \pi_j \pi_k \\
        \dot{\pi}_i &= F_{ij} \dot{x}_j - e E_i + \mathcal{O}(\lambda^{-4}_s).
    \end{aligned}
\end{equation}
By solving the Boltzmann equation up to second order in the electric field and third order in spatial gradients, and evaluating the current using the resulting distribution function, we find that the nonreciprocal conductivity $\sigma_{ijk}$ is given by (see detailed derivations in Appendix \ref{appendix_nonreciprocal_geodesic})
\begin{equation} \label{eq_cond_nr}
    \sigma_{ijk} = \frac{3 \tau^2 e^3 n}{2m^3J^2} \braket{\Gamma_{(ijk)}}.
\end{equation}
Thus, the nonreciprocal conductivity is governed by the spatial average of the geodesic SSC.

This nonreciprocal transport is a type of emergent response induced by the spin texture, similar to the THE described by Eq.~(\ref{eq_cond_H}).
There are several similarities as well as differences between the two.
We first discuss the similarities and differences in the underlying magnetic structures.
Both effects are finite only when the magnetic structure is noncoplanar.
The THE becomes finite when the conventional SSC associated with a finite solid angle is nonzero and has a nonzero spatial average in $\mathbb{R}^{d=2,3}$. 
In particular, for $d=2$, this condition corresponds to a finite skyrmion number $N_{\mathrm{sk}}$, namely a topologically nontrivial spin texture.
In contrast, for the emergent nonreciprocal transport, 
it suffices that the geodesic SSC $\Gamma_{ijk}$ and its spatial average are finite, exemplified by conical magnets, and they can be finite even in one-dimensional systems with $d=1$ or $d_{\mathrm{p}}=1$.
In addition, the system does not need to be topologically nontrivial.

Next, we discuss the differences in the approximation.
Since the THE is independent of the exchange coupling constant $J$, it is an effect obtained within the adiabatic approximation.
In contrast, the emergent nonreciprocal transport depends on $J^{-2}$, which corresponds to the second-order nonadiabatic effect.
Furthermore, while the SSC $\chi_{ij}$ contributes at second order in gradients ($\lambda_s^{-2}$), the emergent nonreciprocal transport appears at third order in gradients ($\lambda_s^{-3}$), i.e., it is higher by one order in the gradient expansion.
Since the contribution to nonreciprocal transport involves higher-order corrections
beyond the conventional semiclassical theory formulated within the adiabatic approximation, we might expect it to be negligibly small.
However, as we will demonstrate with explicit examples below, the nonreciprocal conductivity is not negligibly small.

Finally, we discuss the similarities and differences in the underlying mechanisms.
In the derivations presented here, spin-orbit coupling (SOC) is not introduced into the system.
Therefore, both the THE and the emergent nonreciprocal transport are SOC-free responses.
In both cases, the spin texture directly influences the orbital motion of electrons.
As can be seen from the semiclassical equations of motion, the SSC acts as an effective orbital magnetic field, as expressed in Eq.~(\ref{eq_canonical_1}), thereby generating a Hall effect.
In contrast, as shown in Eq.~(\ref{eq_Hamiltonian_Gamma_asymmetry}), the geodesic SSC effectively induces band asymmetry, leading to an asymmetry between right-moving and left-moving orbital motions, which gives rise to the nonreciprocal transport.
These differences discussed above are summarized in Table~\ref{table_responses}.

As shown in Eq.~(\ref{eq_cond_nr}), the nonreciprocal conductivity is described by the geodesic SSC, which is related to the geodesic curvature.
We note that the geodesic curvature has also appeared in momentum-space quantum geometry in the specific context of the shift photocurrent response of two-band systems through the shift vector~\cite{10.21468/SciPostPhys.11.4.075,Wang_2026}. 
In contrast, Eq.~(\ref{eq_cond_nr}) describes a dc nonreciprocal response: the underlying mechanism is the nonlinear Drude weight, and the relevant geodesic curvature originates from the real-space geometry of the spin trajectory on $\mathbb{S}^2$.

Finally, we estimate the order of magnitude of the nonreciprocal conductivity.
For conical magnets, the spatial average of the geodesic SSC is given by $\braket{\Gamma_{xxx}} = \hbar^3 q^3 \cos^2 \alpha \sin \alpha /4$ from Eq.~(\ref{eq_gamma_conical}), and therefore, the nonreciprocal conductivity is
\begin{equation} \label{eq_NRR_conical}
    \sigma_{xxx} = \frac{3 \hbar^3 \tau^2 e^3 n q^3 \cos^2 \alpha \sin \alpha}{8m^3 J^2}.
\end{equation}
The nonreciprocal transport has been studied in conical magnets in several experimental and theoretical works \cite{doi:10.1126/sciadv.adw8023,doi:10.1073/pnas.2405839122,ltwf-zhj2,jiang2020electric,PhysRevLett.122.057206,PhysRevB.103.184428,https://doi.org/10.1002/adma.202420614,lhj4-9h29}.
In theoretical analyses of the nonreciprocal conductivity originating from the conical spin texture in Ref.~\cite{doi:10.1126/sciadv.adw8023}, a closed formulation in momentum space has been employed to compute the conductivity associated with the nonlinear Drude weight.
We find that the resulting expression is in complete agreement with Eq.~(\ref{eq_NRR_conical}), including the numerical factors, although there is no reason for them to match.
If we set $m=9.1 \times 10^{-31}~\mathrm{kg}$, $J=0.1~\mathrm{eV}$, $n=10~\mathrm{nm}^{-3}$, $q=0.1~\mathrm{nm}^{-1}$, and $\tau = 10~\mathrm{fs}$, the nonreciprocal conductivity is estimated to be $3.6 \times10^{-6}~\mathrm{A/V^2}$.
For the two-dimensional skyrmion crystal discussed in Sec.~\ref{sec_two_dim}, we obtain $\sigma_{ijk} =  Q_0^3 \times 10^{-3}~\mathrm{A/V^2}$, using $\braket{\gamma_{ijk}} \approx 0.1 Q^3 = 0.1 Q_0^3~\mathrm{nm^{-3}}$ and $Q = Q_0~\mathrm{nm}^{-1}$.
For $Q_0 = 0.1$, the result is of the same order as that for conical magnetism.
The magnitude of the nonreciprocal conductivity can be substantially enhanced by considering spin textures with shorter periods, electrons with lighter effective masses, cleaner metals, or exchange interactions closer to the weak-coupling regime.
In addition to these examples, there are some theoretical studies on the nonreciprocal responses in other noncoplanar magnetic systems \cite{doi:10.7566/JPSJ.91.094704,PhysRevB.106.014420,solids6030040}, which are also expected to be understood in terms of the mechanism of the geodesic SSC.

\section{Conclusion, Discussion, and Outlook}
In this work, we have provided a classification of magnetic textures into collinear, coplanar, and noncoplanar magnets based on the differential geometry of curves and surfaces.
In magnetic systems, spins in real space trace curves and surfaces on the unit sphere, which can be characterized by several geometric quantities. We utilize these geometric quantities to classify magnetic textures.
There are four key geometric quantities: the Riemannian metric, the Berry curvature (the infinitesimal area element), the geodesic curvature, and the torsion. 
These quantities are directly related to spin chiralities, namely, the vector spin chirality, the scalar spin chirality, the geodesic scalar spin chirality, and the torsional scalar spin chirality, respectively. 
The first captures noncollinearity, whereas the latter three characterize noncoplanarity.
We obtain five classes of magnetic textures up to the parameter dimension $d_{\mathrm{p}} = 2$, depending on whether these spin chiralities vanish or remain finite, as summarized in Table~\ref{table_classification}. 
In particular, we refine the classification of noncoplanar magnets and identify three distinct types.
Type-I and Type-II are new classes characterized by a vanishing conventional scalar spin chirality (SSC) but a finite geodesic SSC or torsional SSC, providing a complete classification for magnets with $d_{\mathrm{p}} = 1$.
A notable point is that these new spin chiralities remain finite even in magnetic systems with $d_{\mathrm{p}} = 1$, where the conventional SSC necessarily vanishes, leading to an incomplete characterization of noncoplanar magnets in the conventional framework. 
Conical magnets are representative examples belonging to these new noncoplanar classes, although they have been previously ambiguous in the conventional classification.
Moreover, the geodesic SSC is useful for characterizing noncentrosymmetric magnets. 
Its spatial average can serve as an order parameter with the same symmetry as the magnetic toroidal dipole, magnetic quadrupole, and magnetic toroidal octupole. 
Importantly, the geodesic SSC itself represents the corresponding multipole densities and is independent of the choice of the real-space origin.

Furthermore, we have revealed novel emergent phenomena induced by spin chiralities beyond the topological Hall effect associated with the conventional SSC, as summarized in Table~\ref{table_responses}. 
By extending the semiclassical theory to include nonadiabatic effects and higher-order gradients, we find that the Riemannian metric generates a repulsive potential and an effective mass enhancement, leading to changes in resistivity. 
Since the Riemannian metric can be interpreted as the exchange energy of spins, sharp resistivity changes are expected near topological transitions.
In addition, we find that the geodesic SSC appears directly in the semiclassical Hamiltonian, giving rise to an emergent band asymmetry. 
This band asymmetry, in turn, produces a nonreciprocal response, providing a general mechanism originating from magnetic textures. 
These emergent responses, including the topological Hall effect, are spin-orbit-coupling-free and purely orbital effects induced by magnetic textures.

Here, we discuss the experimental aspects of the emergent responses.
The electrical resistivity associated with the Riemannian metric is usually challenging to isolate experimentally, since the Drude contribution is typically dominant.  
However, recent experiments have observed a large jump in the electrical resistivity in the skyrmion phase of skyrmion systems with relatively small diameters, which is expected to originate from nonadiabatic effects \cite{13yj-sbgc,khanh2020nanometric,takagi2022square,Yoshimochi2024Multistep}. 
Therefore, to clarify its origin, it is important to examine the electrical resistivity originating from magnetic textures in such systems, which remains an important subject for future study.

Regarding the nonreciprocal response, it has been experimentally observed in various magnetic materials, and several mechanisms have been proposed \cite{doi:10.1126/sciadv.aat1115,PhysRevMaterials.8.044407,yokouchi2017electrical,PhysRevB.103.L220410,PhysRevB.103.184428,PhysRevLett.122.057206,doi:10.1073/pnas.2405839122,doi:10.1126/sciadv.adw8023,ltwf-zhj2,jiang2020electric,https://doi.org/10.1002/adma.202420614,doi:10.1126/sciadv.abd3703}.
These include a nonreciprocal scattering mechanism dominated by magnetic fluctuations associated with spin clusters \cite{ishizuka2020anomalous,PhysRevMaterials.8.044407,yokouchi2017electrical,PhysRevB.103.L220410,ltwf-zhj2}, a mechanism based on current-induced deformation of the magnetic structure \cite{doi:10.1126/sciadv.aat1115}, and band asymmetry induced by the magnetic texture \cite{doi:10.1073/pnas.2405839122,doi:10.1126/sciadv.adw8023,https://doi.org/10.1002/adma.202420614,lhj4-9h29}.
In this work, we have focused on the last one.
In these materials, the magnetic textures have a finite geodesic SSC; thus, the band asymmetry and the resulting nonreciprocal response are expected to originate from the geodesic SSC.
Here, we estimate the nonreciprocal conductivity in Eq.~(\ref{eq_NRR_conical}) for materials in which the nonreciprocal conductivity originating from the mechanism has been measured.
As a first example, the centrosymmetric hexagonal crystal $\mathrm{YMn_6Sn_6}$ hosts a helical magnetic phase that breaks spatial inversion symmetry.
In this material, a magnetic-field-induced transverse conical state emerges. 
Nonreciprocal transport has been observed in this phase, with an enhanced response near the phase boundary \cite{https://doi.org/10.1002/adma.202420614}.
While the enhanced response near the phase boundary has been attributed to nonreciprocal scattering by spin clusters, the nonreciprocal response within the transverse conical phase is considered to originate from band asymmetry. 
In this material, the Mn $3d$ electrons are responsible for its metallic magnetism, which is described within a Hund's-metal picture. 
For the material parameters, we use a helical pitch of $3~\mathrm{nm}$, a carrier density of $n \sim 1.0 \times 10^{22}~\mathrm{cm}^{-3}$, an effective mass of $m^* \sim m$, a Hund's coupling of $J \sim 0.5~\mathrm{eV}$, and a relaxation time of $\tau \sim 0.1~\mathrm{ps}$ \cite{li2021dirac,PhysRevB.103.014416}.
Then, the nonreciprocal conductivity is $\sigma_{xxx} \sim 1.3 \times 10^{-1}~\mathrm{A/V^2}$.
In experiments, the nonreciprocal resistivity (NRR) has been measured.
The NRR is estimated as 
$\mathrm{NRR} = \sigma_{xxx}\rho_{xx}^{3}j$.
Here, $\rho_{xx} \sim 10~\mathrm{\mu \Omega\,cm}$ is the longitudinal resistivity, and $j = 5\times10^8~\mathrm{Am^{-2}}$ is the applied current density.
Using this relation, we obtain $\mathrm{NRR} \sim 6.6 \times 10^{-3}~\mathrm{n\Omega\,cm}$. 
This value differs from the experimental value, $10^{-1}~\mathrm{n\Omega\,cm}$, by roughly one to two orders of magnitude.
As another example, $\alpha$-EuP$_3$ is a centrosymmetric semimetal in which nonreciprocal transport has been observed in the field-induced conical magnetic phase \cite{doi:10.1073/pnas.2405839122,PhysRevX.12.011033}. 
The localized magnetic moments originate from the Eu-$f$ electrons, while the conduction electrons at the Fermi surface are composed of Eu-$s$ and P-$p$ orbitals. The Hund's coupling between the Eu-$f$ and Eu-$s$ electrons is estimated to be on the order of $J \sim 0.1~\mathrm{eV}$.
This material exhibits helical magnetic order with a wave number of approximately 
$q = 5.6~\mathrm{nm}^{-1}$. The carrier density, effective mass, and relaxation time are estimated as 
$n \sim 1.0 \times 10^{19}~\mathrm{cm}^{-3}$, 
$m^* \sim 0.1m$, and 
$\tau \sim 10~\mathrm{fs}$, respectively.
These parameters give $\sigma_{xxx} \sim 6.3 \times 10^{-1}~\mathrm{A/V^2}$. 
Using $\rho_{xx} \sim 1~\mathrm{m\Omega\,cm}$ and 
$j=1\times 10^{7}~\mathrm{A/m^2}$, we obtain 
$\mathrm{NRR} \sim 0.63~\mathrm{\mu\Omega\,cm}$, which roughly coincides with the experimental value.
As another example, band-asymmetry-driven nonreciprocal transport has also been observed in $\mathrm{Co_8Zn_9Mn_3}$ \cite{doi:10.1126/sciadv.adw8023}. 
As discussed in the main text, our expression in Eq.~(\ref{eq_NRR_conical}) is identical, including its coefficient, to the equation used in the theoretical analysis of this material. 
These estimations demonstrate that our formula can provide a directly relevant basis for comparison with experiment.

Although the nonreciprocal transport arising from band asymmetry in magnetic materials has thus been reported experimentally, our theory goes one step further by identifying its origin as the geodesic SSC. This is also advantageous from an experimental viewpoint. The geodesic SSC can be extracted once the magnetic texture is determined, for example by neutron scattering or, more directly, by Lorentz microscopy. Therefore, the nonreciprocal conductivity and the geodesic SSC can be compared using experimental data alone, providing a more direct way to identify the origin of nonreciprocal transport.

We comment on the assumptions adopted in this work and on regimes beyond them.
We have considered a simple model in which a free-electron gas is coupled to ordered classical spins, focusing on the strong-coupling limit of this interaction.
This setting is relevant, for example, to systems in which localized spins and conduction-electron spins are coupled through a Hund's coupling, as in double-exchange models, or to mean-field descriptions of itinerant magnets with large magnetic moments. 
Since our analysis also relies on a gradient expansion, we have further assumed that the spatial modulation of the magnetic texture is sufficiently smooth.
While these conditions are commonly satisfied, they do not hold in all cases.
In real materials, there are also systems in the weak-coupling regime, as well as magnetic lattice systems with small magnetic unit cells, for which the gradient expansion may not be quantitatively valid. 
Extending the present framework to such systems is therefore an important direction for future work. 
Nevertheless, in the topological Hall effect described by conventional emergent electromagnetism, the scalar-spin-chirality picture remains essential even in such regimes \cite{PhysRevB.92.115417,PhysRevB.62.R6065,doi:10.1143/JPSJ.71.2613}. 
By analogy, we expect that the picture of the emergent nonreciprocal transport based on the geodesic SSC will also play an important role in these broader classes of systems.
In addition, we have not included effects arising from spin-orbit coupling $\lambda_{\mathrm{SOC}}$. 
Such effects are expected to enter only as corrections of relative order $\lambda_{\mathrm{SOC}}/J$ and therefore constitute subleading contributions in the strong-coupling regime considered here. 
Indeed, the topological Hall effect is almost insensitive to spin-orbit coupling \cite{doi:10.1126/sciadv.abq2765}. 
We expect a similar robustness for the emergent nonreciprocal transport discussed in this work.

Furthermore, we have formulated the classification and emergent responses in the high-symmetry continuum limit, where the orbital space is treated as a uniform isotropic space with O(3) symmetry. 
Strictly speaking, in real materials, the orbital space is discretized and described by lower-symmetry crystalline groups.
In such cases, the geodesic SSC should also be discretized and formulated in the language of discrete differential geometry.
In addition, to identify a broader class of materials in which the discrete geodesic SSC becomes finite and gives rise to SOC-free emergent phenomena, it will be an important direction for future work to extend the present classification by incorporating spin crystallographic groups \cite{PhysRevB.109.094438}.
On the other hand, in systems with SOC, emergent responses do not necessarily require noncoplanar magnetic textures. 
For the conventional SSC, SOC-modified forms of SSC have been discussed and provided a basis for understanding recent developments such as the anomalous Hall effect in fully compensated antiferromagnets \cite{PhysRevB.101.024420,PhysRevB.102.075112,nakatsuji2015large}. 
In a similar spirit, it will be important to clarify how SOC-induced corrections enter the geodesic SSC, and thereby to develop a unified understanding of emergent phenomena in a broader class of materials.

Finally, we expect that the present classification based on differential geometry will serve as a foundation for further exploring the classification and characterization of other orders, beyond topology and symmetry, and not limited to magnetic orders.
More generally, the manifolds traced by such order parameters can be more complex, giving rise to a richer variety of geometric structures and more diverse classes.
Indeed, the dimensionality of order parameters is not necessarily restricted to two and can be higher. Examples include spin-density-wave states with amplitude modulation, spin-triplet superconducting states, multiferroic magnets characterized by coupled order parameters such as polarization and magnetization, and more general multipolar orders described by higher-rank tensor order parameters.
Furthermore, the present formulation of semiclassical theory, extended to incorporate nonadiabatic effects and higher-order gradients, can be applied to other systems coupled to electrons and is expected to produce novel quantum geometric effects beyond the Berry curvature and the quantum metric.
In this sense, our framework is expected to further promote emergent electrodynamics and a broader class of emergent phenomena in condensed matter physics.

\section*{ACKNOWLEDGMENTS}
We acknowledge E. Barts and W. Koshibae for useful discussions.
K.S. acknowledges financial support from the RIKEN Special Postdoctoral Researcher (SPDR) Program. N.N. was supported by JSPS KAKENHI Grants No. 24H00197, No.24H02231, and No. 24K00583. N.N. was supported by the RIKEN TRIP initiative.

\appendix

\section{Derivation of the torsional scalar spin chirality} \label{appendix_torsion}
We consider a curve $\bm{p}(w)$ in three-dimensional space.
The following discussion follows standard treatments in textbooks on differential geometry \cite{Kobayashi2019DGCS}.
We introduce the Frenet frame ${\bm{e}_1, \bm{e}_2, \bm{e}_3}$, which forms an orthonormal basis ($\bm{e}_i \cdot \bm{e}_j = \delta_{ij}$) defined using the arc-length parameter $s$ as follows:
\begin{equation}
    \bm{e}_1 = \bm{p}'(s),~~\bm{e}'_1 = \kappa \bm{e}_2,~~\bm{e}_3 = \bm{e}_1 \times \bm{e}_2,
\end{equation}
where $\kappa = |\bm{e}'_1|$ denotes the curvature of the curve $\bm{p}$.
In the following, a prime means differentiation with respect to the arc-length parameter $s$.
The Frenet-Serret formulas for a space curve are given by
\begin{equation}
    \begin{pmatrix}
        \bm{e}'_1 \\
        \bm{e}'_2 \\
        \bm{e}'_3
    \end{pmatrix}
    =
    \begin{pmatrix}
        0 & \kappa & 0 \\
        -\kappa & 0 & \tau \\
        0 & -\tau & 0 \\
    \end{pmatrix}
    \begin{pmatrix}
        \bm{e}_1 \\
        \bm{e}_2 \\
        \bm{e}_3 
    \end{pmatrix},
\end{equation}
where $\tau = \bm{e}_3 \cdot \bm{e}'_2 = -\bm{e}'_3 \cdot \bm{e}_2$ is the torsion of the curve.
We obtain an identity using the Frenet-Serret formulas as
\begin{align} \label{eq_tau_kappa}
    \tau &= \bm{e}_3 \cdot \bm{e}'_2 \nonumber \\
    &= (\bm{e}_1 \times \bm{e}_2) \cdot \kappa^{-1} \bm{e}''_1.
\end{align}

We define a general torsional scalar spin chirality for the curve $\bm{p}$ by
\begin{equation}
    t[\bm{p}] = \partial_w\bm{p} \cdot ( \partial_{ww} \bm{p} \times \partial_{www} \bm{p} ).
\end{equation}
When expressed in terms of the arc-length parameter, the general torsional scalar spin chirality is given by
\begin{align}
    t[\bm{p}] &= | \partial_w \bm{p}|^6 \bm{p}' \cdot ( \bm{p}''  \times \bm{p}''' ) \nonumber \\
    &= | \partial_w \bm{p}|^6   \bm{e}_1 \cdot ( \kappa \bm{e}_2 \times \bm{e}''_1).
\end{align}
Using the relation in Eq.~(\ref{eq_tau_kappa}), the torsional scalar spin chirality is written as
\begin{equation}
    t[\bm{p}] = \tau  \kappa^2 |\partial_w \bm{p}|^6.
\end{equation}

In particular, when restricted to a curve on the unit sphere ($\bm{p} = \bm{n},~|\bm{n}|=1$), $\bm{n}'$ and $\bm{n}''$ are given by
\begin{equation}
\begin{aligned}
    &\bm{n}' = \bm{e}_1 \\
    &\bm{n}'' = \kappa \bm{e}_2 = - \bm{n} + \kappa_g (\bm{n} \times \bm{n}').
\end{aligned}
\end{equation}
Thus, the curvature reads $\kappa = \sqrt{1 + \kappa^2_g}$.
Furthermore, the torsion $\tau$ can be expressed in terms of the curvature and the geodesic curvature.
$\bm{e}_3$ is given by
\begin{equation}
    \bm{e}_3 
    = 
    \bm{e}_1 \times \bm{e}_2  
    =
    \frac{1}{\kappa} ( \kappa_g \bm{n} + \bm{n} \times \bm{n}' ),
\end{equation}
and $\bm{e}'_2$ is given by
\begin{equation}
    \bm{e}'_2 = -\frac{\kappa'}{\kappa} \bm{e}_2 + \frac{1}{\kappa} ( - (1 + \kappa_g) \bm{n}' + \kappa_g' ( \bm{n} \times \bm{n}' )).
\end{equation}
Therefore, we can express the torsion as follows:
\begin{equation}
    \tau= \bm{e}_3 \cdot \bm{e}'_2 = \frac{\kappa_g'}{\kappa^2} = \frac{\kappa_g'}{1 + \kappa_g^2}.
\end{equation}

\section{Derivation of the quantum effective low-energy Hamiltonian} \label{appendix_quantum_ham}
The Hamiltonian in the local spin frame reads
\begin{align}
    \hat{H}' &= U^\dagger(\hat{\bm{x}}) \hat{H} U(\hat{\bm{x}}) \nonumber \\
    &= \frac{1}{2m} (\hat{\bm{p}} + e \bm{A}(t) - \bm{a}(\hat{\bm{x}}))^2 - J \sigma_z.
\end{align}
Here, $\bm{a}(\hat{\bm{x}}) = i U \bm{\nabla} U$ is a pure SU(2) gauge field.
In the following, we regard $J$ as the largest energy scale and start from the lowest-energy state in the $-J$ sector, incorporating corrections perturbatively in powers of $J^{-1}$.
Because the off-diagonal components of the SU(2) gauge field induce mixing between the $\pm J$ sectors, we compute the low-energy effective Hamiltonian using $J^{-1}$ as the perturbative expansion parameter.
The kinetic Hamiltonian is decomposed into the diagonal part and the off-diagonal part as (hereafter, we omit the hat notation for operators.)
\begin{align}
    &\begin{pmatrix}
        \bm{\pi}_1 & - \bm{a}_{12} \\
         - \bm{a}^*_{12}  &  \bm{\pi}_2
    \end{pmatrix}^2 \nonumber\\
    &=
    \begin{pmatrix}
        \bm{\pi}_1^2 + | \bm{a}_{12} |^2
        & 0 \\
        0 & \bm{\pi}_2^2 + | \bm{a}_{12}|^2
    \end{pmatrix} \nonumber \\
    &\quad+
    \begin{pmatrix}
        0 & - \{ \bm{\pi} , \bm{a}_{12} \} \\
        - \{ \bm{\pi} , \bm{a}_{12}^* \} & 0
    \end{pmatrix}.
\end{align}
Here, we use $\bm{\pi} = \bm{p} + e\bm{A}$, $\bm{\pi}_i = \bm{\pi} - \bm{a}_{ii} = \bm{\pi}  \mp \bm{a}_z , \bm{a}_z = ( \bm{a}_{11} - \bm{a}_{22})/2 $.
Then, the total Hamiltonian is given by
\begin{gather}
    H'
    = H_0 + H_K  + V, \nonumber \\
    H_0 =
    -J\sigma_z
    ,~
    H_K =
    \begin{pmatrix}
        H_1 & 0\\
        0 & H_2 
    \end{pmatrix},
    ~
    V=
    \begin{pmatrix}
        0 & \Lambda \\
        \Lambda^\dagger & 0
    \end{pmatrix},
\end{gather}
where $H_0 + H_K$ is the diagonal part and $V$ is the off-diagonal part.
Here, we define $H_i = \frac{1}{2m} ( \bm{\pi}_i^2 + |\bm{a}_{12}|^2 )$ and $\Lambda = - \frac{1}{2m} \{ \pi_i , a_{12i}  \}$.

Next, we systematically perform the expansion in powers of $J^{-1}$ using the Schrieffer-Wolff (SW) transformation \cite{PhysRev.149.491}.
Using a unitary matrix $e^S$, we perform a unitary transformation on $H'$ as
\begin{align}
    H'' &= e^S H' e^{-S} \nonumber \\
    &=
    H' + [ S, H'] + \frac{1}{2} [ S ,[ S, H' ]] + \frac{1}{6}[S, [S, [S, H']]] + \cdots.
\end{align}
We determine the generator $S$ so as to eliminate the off-diagonal terms, imposing this condition order by order in powers of $J^{-a}$.
Accordingly, we expand
\begin{equation}
S = S^{(1)} + S^{(2)} + \cdots ,
\end{equation}
where each $S^{(a)}$ is of order $J^{-a}$ and is fixed independently such that the off-diagonal terms vanish at the corresponding order.
Then, we obtain the expansion
\begin{widetext}
\begin{align}
    H''
    &=
    \underbrace{H_0}_{\propto \mathcal{O}(J)} + \underbrace{H_K + V + [ S^{(1)} , H_0 ]}_{\propto \mathcal{O}(J^0)}
    +
    \underbrace{[ S^{(1)} , H_K + V ] + \frac{1}{2} [ S^{(1)} , [ S^{(1)} , H_0  ]]
    + [ S^{(2)} , H_0 ]}_{\propto \mathcal{O}( J^{-1} )} \nonumber \\
    &\quad +\underbrace{
    \frac{1}{2} [ S^{(1)} , [ S^{(1)} , H_K + V  ]]
    +[ S^{(2)} , H_K + V ] 
    + \frac{1}{2}  [ S^{(1)} , [ S^{(2)} , H_0 ]]
    + \frac{1}{2}  [ S^{(2)} , [ S^{(1)} , H_0 ]]
    +
    \frac{1}{6}[S^{(1)}, [S^{(1)}, [S^{(1)}, H_0]]]
    }_{\propto \mathcal{O}( J^{-2} )} \nonumber \\
    &\quad + \mathcal{O}(J^{-3}).
\end{align}
\end{widetext}

\subsection{First-order nonadiabatic effect}
We first determine $S^{(1)}$ so as to eliminate the off-diagonal terms at order $J^{0}$ \cite{qxnw-8q4y}.
The generator $S^{(1)}$ satisfies the relation
\begin{equation}
    V + [ S^{(1)} , H_0 ] = 0.
\end{equation}
The generator $S^{(1)}$ that satisfies this relation is given by
\begin{equation}
    S^{(1)}_{11} = S^{(1)}_{22} = 0,~~~
    S^{(1)}_{12} = - S^{(1)\dagger}_{21} = -\frac{\Lambda}{2J}.
\end{equation}
Using this result, we evaluate the diagonal part at order $J^{-1}$.
Note that $S^{(2)}$ is introduced to cancel the off-diagonal terms at this order and therefore does not contribute to the diagonal sector.
Furthermore, $[S^{(1)}, H_K]$ also generates only off-diagonal contributions.
The diagonal part at order $J^{-1}$ is 
\begin{align}
    H^{\mathrm{diag}}_{\mathrm{nad,1}}
    &=[S^{(1)} , V ] + \frac{1}{2} [S^{(1)} , [  S^{(1)} , H_0 ]] = \frac{1}{2} [ S^{(1)} , V] \nonumber \\
    &=
    \frac{1}{2J}
    \begin{pmatrix}
        - \Lambda \Lambda^\dagger & 0 \\
        0 & \Lambda \Lambda^\dagger \\
    \end{pmatrix}.
\end{align}

\subsection{Second-order nonadiabatic effect}
We determine $S^{(2)}$ so as to eliminate the off-diagonal terms at order $J^{-1}$.
The generator $S^{(2)}$ satisfies the relation
\begin{equation}
    [ S^{(1)} , H_K ] + [ S^{(2)} , H_0 ] = 0.
\end{equation}
Solving the relation, we obtain
\begin{equation}
\begin{aligned}
    &S^{(1)}_{11} = S^{(1)}_{22} = 0, \\
    &S^{(2)}_{12} = - S^{(2)\dagger}_{21} = \frac{1}{4J^2} ( H_1 \Lambda - \Lambda H_2  ).
\end{aligned}
\end{equation}
Using the obtained generator $S^{(2)}$, the diagonal part at order $J^{-2}$ is
\begin{align}
    &\frac{1}{2}[ S^{(1)}, [ S^{(1)}, H_K ] ] + [ S^{(2)} ,V ] \nonumber \\
    &+ \frac{1}{2}[ S^{(1)} , [ S^{(2)} , H_0 ] ] + \frac{1}{2}[ S^{(2)} , [ S^{(1)} , H_0 ] ] \nonumber \\
    &= \frac{1}{2} [ S^{(2)}, V ].
\end{align}
In particular, extracting the $(1,1)$ component, we obtain
\begin{equation}
    \frac{1}{2}[ S^{(2)} , V ]_{11}
    =
    \frac{1}{8J^2} ( \{ H_1 , \Lambda \Lambda^\dagger \} - 2\Lambda H_2 \Lambda^\dagger ).
\end{equation}

\vskip\baselineskip
Finally, we obtain the quantum effective low energy Hamiltonian consisting of
\begin{equation}
    \begin{aligned}
        &H_{\mathrm{ad}} = -J + \frac{1}{2m} (\bm{\pi}_{1}^2 + |\bm{a}_{12}|^2), \\
        &H_{\mathrm{nad},1} = - \frac{\Lambda \Lambda^\dagger}{2J}, \\
        &H_{\mathrm{nad},2} = \frac{1}{8J^2} (  \{ H_1, \Lambda \Lambda^{\dagger} \} - 2 \Lambda H_2 \Lambda^{\dagger} ),
    \end{aligned}
\end{equation}
by extracting the $-J$ sector.

\section{Derivation of the semiclassical Hamiltonian} \label{appendix_semi_ham}
We apply the Wigner transformation to the quantum effective low energy Hamiltonian and map it onto its classical counterpart in phase space.
The Wigner transformation is defined by
\begin{equation}
    A_W(\bm{x} , \bm{p}) = \int d^dy e^{ -i \bm{p} \cdot \bm{y}} \braket{ \bm{x}+\bm{y}/2 | \hat{A} | \bm{x} - \bm{y}/2  }.
\end{equation}
Here, $(\bm{x},\bm{p})$ are the canonical variables.
Under the Wigner transformation, the product of operators is replaced by the Moyal product as
\begin{equation}
    \hat{A} \hat{B} \to A_W \star B_W,~~~\star = \exp \Bigl(\frac{i \hbar}{2} ( \overleftarrow{\partial_{\bm{x}}} \cdot \overrightarrow{\partial_{\bm{p}}} - \overleftarrow{\partial_{\bm{p}}} \cdot \overrightarrow{\partial_{\bm{x}}}  )  \Bigr).
\end{equation}
One important property of the Moyal product is that, for an operator $\hat{A}$ that depends at most linearly on $\hat{\bm{x}}$ or $\hat{\bm{p}}$, the Moyal product appearing in the squared operator $\hat{A}^2$ reduces to the ordinary product as
\begin{equation} \label{eq_AA}
    \hat{A} \hat{A} \to (A_W)^2~~~( \hat{A} = \mathcal{O}(\hat{\bm{x}}^1)~\text{or}~ \mathcal{O}(\hat{\bm{p}}^1)).
\end{equation}
Since the operator depends at most linearly on $\hat{\bm{x}}$ or $\hat{\bm{p}}$, all terms of second and higher order in the gradient (or $\hbar$) expansion vanish.
The first-order contribution originates from the noncommutativity of operators; however, in the present case, it also vanishes because the product involves identical operators and is therefore commutative.

First, we perform the Wigner transformation of the Hamiltonian within the adiabatic approximation $\hat{H}_{\mathrm{ad}}$.
Since the Hamiltonian satisfies the property in Eq.~(\ref{eq_AA}), the procedure reduces to replacing operators with their classical counterparts.
Then, we obtain the semiclassical Hamiltonian within the adiabatic approximation as
\begin{equation}
    H_{\mathrm{ad}}^{\mathrm{cl}} = \frac{1}{2m} ( \bm{\pi}_1^2 + |\bm{a}_{12}|^2 ).
\end{equation}

Next, we perform the Wigner transformation of the Hamiltonian arising from nonadiabatic effects at order $J^{-1}$.
Since $\hat{\Lambda} = -\frac{1}{2m} \{ \hat{\pi}_i , a_{12i} \}$ satisfies the property in Eq.~(\ref{eq_AA}), the Wigner transformation yields
\begin{equation}
    \{ \hat{\pi}_i , a_{12i} \} \to \{ \pi_i , a_{12i} \}.
\end{equation}
Using this expression, we obtain
\begin{align}
    &\{ \hat{\pi}_i , a_{12i} \} \{a_{12j}^* , \hat{\pi}_j   \} \nonumber \\
    &\to
    4 a_{12i} a^{*}_{12j} \pi_i \pi_j
    + 4 \mathrm{Im} [ a_{12i} \partial_i a^{*}_{12j}  ] \pi_j \nonumber \\
    &\quad+ \partial_i a_{12j} \partial_j a^{*}_{12i}.
\end{align}
In contrast to the adiabatic approximation, it is noteworthy that the term originating from the nonadiabatic effect gives rise to higher-order contributions in the gradient expansion of the Moyal product, namely contributions of order $\hbar^1$ and beyond.
Then, we obtain the semiclassical Hamiltonian at order $J^{-1}$ as
\begin{align}
    H_{\mathrm{nad,1}}^{\mathrm{cl}}
    =&
    -
    \frac{1}{2m^2J} \Bigl( a_{12i} a_{12j}^{*} \pi_i \pi_j + \mathrm{Im}[ a_{12i} \partial_i a_{12j}^{*}  ] \pi_j  \nonumber \\
    &+ \frac{1}{4} \partial_j a_{12i} \partial_i a^{*}_{12j} \Bigr).
\end{align}
Here, three terms appear; however, the coefficients of the momentum $\bm{\pi}$ in each term are not individually gauge invariant.
To resolve this, we rewrite the momentum using $\bm{\pi} = \bm{\pi}_1 + \bm{a}_z$, thereby expressing all terms in terms of $\bm{\pi}_1$.
As a result, we obtain a Hamiltonian in which each term is independently gauge invariant, given by
\begin{align}
    H_{\mathrm{nad},1}^{\mathrm{cl}}
    =&
    -
    \frac{1}{2m^2J} \Bigl( \mathrm{Re} [a_{12i} a^{*}_{12j}] \pi_{1i} \pi_{1j} 
    + \mathrm{Im} [ a_{12i} D_i a^{*}_{12j}  ] \pi_{1j} \nonumber \\
    &+ \frac{1}{4} \mathrm{Re}[ D_j^* a_{12i} D_i a^{*}_{12j} ]
    \Bigr) \nonumber \\
    &=
     -\frac{1}{2m^2 J} ( G_{ij} \pi_{1i} \pi_{1j} + \Gamma_i \pi_{1i} + R ).
\end{align}
Here, $D_i = \partial_i + 2i a_{zi}$ is the covariant derivative.

The Hamiltonian of the second-order nonadiabatic correction is also Wigner-transformed in the same manner.
In the following, we retain terms up to third order in spatial gradients $\mathcal{O}(\lambda_s^{-3})$.
Performing the Wigner transformation of the first term in $H_{\mathrm{nad},2}$, we obtain
\begin{align}
    \frac{1}{8J^2} \{ \hat{H}_1 , \hat{\Lambda} \hat{\Lambda}^\dagger \} &\to \frac{1}{4J^2} (H_1)_W (\Lambda \Lambda^\dagger)_W + \mathcal{O}(\lambda_s^{-4})\nonumber \\
    &\simeq \frac{1}{8m^3J^2} \bm{\pi}^2_1 ( G_{ij} \pi_{1i} \pi_{1j} + \Gamma_i \pi_{1i} ).
\end{align}
From the effective commutativity arising from the anticommutator of the operators
$\hat{H}_1$ and $\hat{\Lambda}\hat{\Lambda}^{\dagger}$, the first-order term in the
gradient expansion of the Moyal product vanishes.
Next, we perform the Wigner transformation of the second term in $H_{\mathrm{nad},2}$.
Since $\Lambda^{(\dagger)}$ is already of order $\lambda_s^{-1}$,
it is sufficient to keep $H_2$ up to order $\lambda_s^{-1}$.
The Wigner transformation of $H_2$ is given by
\begin{align}
    (H_2)_W &= \frac{1}{2m} ( \bm{\pi}_2^2 + |\bm{a}_{12}|^2) \nonumber \\
    &=
    \frac{1}{2m} ( \bm{\pi}_1^2 + 4 a_{zi} \pi_{1i} ) + \mathcal{O}(\lambda^{-2}_s).
\end{align}
Here, we use $\bm{\pi}_2 = \bm{\pi}_1 + 2 \bm{a}_z$.
Thus, the Wigner transformation of the second term in $H_{\mathrm{nad},2}$ reads
\begin{align}
    -\frac{1}{4J^2}  \hat{\Lambda} \hat{H}_2 \hat{\Lambda}^\dagger \to& - \frac{1}{4J^2} (\Lambda)_W \star (H_2)_W \star (\Lambda^\dagger)_W \nonumber \\
    =& -\frac{1}{8m^3J^2} \Bigl( \pi_i \pi_k a_{12i} a^*_{12k} ( \bm{\pi}_1^2 + 4 a_{zj} \pi_{1j} ) \nonumber \\
    & + (\mathrm{Moyal})\Bigr) + \mathcal{O}(\lambda^{-4}_s)
\end{align}
Here, $(\mathrm{Moyal})$ denotes the contribution arising from the first-order
term of the Moyal product.
Within the order $\lambda_s^{-3}$, this term reads
\begin{align}
    (\mathrm{Moyal})
    =&
    2 \pi_{1i} \pi_{1j} \pi_{1k} \mathrm{Im}[  a_{12i} \partial_k a^{*}_{12j} ] \nonumber \\
    &+ \bm{\pi}_1^2 \pi_{1j} \mathrm{Im}[  a_{12i} \partial_i a^{*}_{12j} ] + \mathcal{O}(\lambda^{-4}_s).
\end{align}
Collecting these results, we obtain
\begin{align}
    &- \frac{1}{4J^2} (\Lambda)_W \star (H_2)_W \star (\Lambda^\dagger)_W \nonumber \\
    &=
    - \frac{1}{8m^3 J^2} \Bigl(
    G_{ij} \pi_{1i} \pi_{1j} \bm{\pi}^2_1 +
        2 \pi_{1i} \pi_{1j} \pi_{1k} \Gamma_{ijk} + \bm{\pi}_1^2 \pi_{1i} \Gamma_i
    \Bigr).
\end{align}
Finally, the correction to the semiclassical Hamiltonian at order $J^{-2}$ is
\begin{equation}
H^{\mathrm{cl}}_{\mathrm{nad},2} = -\frac{1}{4m^3 J^2}\Gamma_{ijk} \pi_{1i} \pi_{1j} \pi_{1k}.
\end{equation}

\section{Spin representation of quantum geometric quantities} \label{appendix_spin_rep}
Here, we present the spin representation of quantum geometric quantities, such as $G_{ij} = \mathrm{Re}[ a_{12i} a^*_{12j}]$ and $\Gamma_{ijk} = \mathrm{Im}[ a_{12i} D_k a^*_{12j} ]$.
The unitary matrix associated with the transformation to the local spin frame is given by
\begin{equation}
    U=
    \frac{1 + ( \bm{n} \cdot \bm{\sigma} ) \sigma_z}{\sqrt{ 2(1 + n_z) }}
    =
    \begin{pmatrix}
        \cos \frac{\theta}{2} & - e^{-i\phi} \sin \frac{\theta}{2} \\
        e^{i \phi } \sin \frac{\theta}{2} & \cos \frac{\theta}{2},
    \end{pmatrix}
\end{equation}
where we represent the spin direction $\bm{n}$ in spherical coordinates as $\bm{n}(\bm{x}) = ( \sin \theta(\bm{x}) \cos \phi(\bm{x}) , \sin \theta(\bm{x}) \sin \phi(\bm{x}) , \cos \theta(\bm{x}))$.
Using this gauge, the Berry connection $\bm{a} = i U^\dagger \bm{\nabla} U$ reads (in the following, we set $\hbar=1$)
\begin{equation}
    \bm{a} = \begin{pmatrix}
        - \frac{1}{2}(1 - \cos \theta)\bm{\nabla} \phi & -\frac{e^{-i\phi}}{2} (  \sin \theta \bm{\nabla} \phi + i \bm{\nabla} \theta ) \\
        -\frac{e^{i\phi}}{2} (  \sin \theta \bm{\nabla} \phi - i \bm{\nabla} \theta ) & \frac{1}{2}(1 - \cos \theta) \bm{\nabla} \phi
    \end{pmatrix}.
\end{equation}
Using this expression, the Riemannian metric $G_{ij}$ is
\begin{equation}
    G_{ij} = \mathrm{Re}[ a_{12i} a^*_{12j}]
    =
    \frac{1}{4} ( \sin^2 \theta \partial_i \phi \partial_j \phi + \partial_i \theta \partial_j \theta ).
\end{equation}
Furthermore, $\partial_i \bm{n} \cdot \partial_j \bm{n}$ can also be expressed in terms of $\theta$ and $\phi$, and by comparing the two expressions, we obtain
\begin{equation}
    G_{ij} = \frac{1}{4} \partial_i \bm{n} \cdot \partial_j \bm{n}.
\end{equation}

Next, we calculate $\Gamma_{ijk} = \mathrm{Im}[ a_{12i} (\partial_k + 2i a_{zk}) a^*_{12j} ]$.
Using the expression for $\partial_k a_{12j}^*$,
\begin{align}
    \partial_k a^{*}_{12j} 
    =&
    -\frac{e^{i\phi}}{2} ( i \sin \theta \partial_j \phi \partial_k \phi  + \partial_j \theta \partial_k \phi \nonumber \\
    &+ \sin \theta \partial_{jk} \phi  + \cos \theta \partial_j \phi \partial_k \theta  - i \partial_{jk} \theta ),
\end{align}
the term $\mathrm{Im}[a_{12i} \partial_k a^{*}_{12j}]$ is given by
\begin{align}
    \mathrm{Im}[a_{12i} \partial_k a^{*}_{12j}]
    =&
    \frac{1}{4} ( \sin^2 \theta \partial_i \phi \partial_j \phi \partial_k \phi - \sin \theta \partial_i \phi \partial_{jk} \theta \nonumber \\
    &+\partial_i \theta \partial_j \theta \partial_k \phi + \sin \theta \partial_i \theta \partial_{jk} \phi \nonumber \\
    &+ \cos \theta \partial_i \theta \partial_j \phi \partial_k \theta 
    ).
\end{align}
On the other hand, the term $\mathrm{Im}[a_{12i} (2i a_{zk}) a^{*}_{12j}]$, using $a_{zk} = (a_{11k}-a_{22k})/2 = -\frac{1}{2} (1 - \cos \theta)\partial_k \phi $, is given by
\begin{align}
    &\mathrm{Im}[a_{12i} (2i a_{zk}) a^{*}_{12j}] = 2a_{zk} G_{ij} \nonumber \\
    &=
    -\frac{1}{4} (1 - \cos \theta) \partial_k \phi (\sin^2 \theta \partial_i \phi \partial_j \phi + \partial_i \theta \partial_j \theta).
\end{align}
Summing these two terms, we obtain the expression for $\Gamma_{ijk}$ as
\begin{align}
    \Gamma_{ijk} =& \frac{1}{4} ( \sin^2 \theta \cos \theta \partial_i \phi \partial_j \phi \partial_k \phi -\sin \theta \partial_i \phi \partial_{jk} \theta \nonumber \\
    &+ \sin \theta \partial_i \theta \partial_{jk} \phi + \cos \theta \partial_i \theta ( \partial_j \theta \partial_k \phi + \partial_j \phi \partial_k \theta  )  ).
\end{align}
This quantity is symmetric under the exchange of the indices $j$ and $k$.
Furthermore, $\bm{n} \cdot ( \partial_i \bm{n} \times \partial_{jk} \bm{n} )$ can be expressed in terms of $\theta$ and $\phi$, and by comparing the two expressions, we obtain
\begin{equation}
    \Gamma_{ijk} = \frac{1}{4} \bm{n} \cdot ( \partial_i \bm{n} \times \partial_{jk} \bm{n} ).
\end{equation}

\section{Boltzmann theory}
\subsection{Resistivity change by the Riemannian metric} \label{appendix_linear_reimann}
We consider a Hamiltonian corrected by the Riemannian metric
\begin{equation}
    H = \frac{\bm{\pi}^2}{2m} - \bar{G}_{ij} \pi_i \pi_j + V,
\end{equation}
where $\bar{G}_{ij} =  G_{ij} / 2m^2 J$ and $V = \mathrm{tr} G / 2m$.
The canonical equations read

\begin{equation} 
    \begin{aligned}
        \dot{x}_i &= \frac{\pi_i}{m} -2 \bar{G}_{ij} \pi_j \equiv f_{xi}, \\
        \dot{\pi}_i &=
        F_{ij} \dot{x}_j -eE_i + \mathcal{O}(\lambda^{-3}_s)
        \equiv f_{\pi i} - e E_i
        .
    \end{aligned}
\end{equation}
The Boltzmann equation is given by
\begin{equation}
    f_{\pi i} \partial_{\pi_i} f + f_{x i} \partial_{x_i} f - e E_i \partial_{\pi_i} f = - \frac{f -f_{\mathrm{eq}}}{\tau}.
\end{equation}
We solve this equation perturbatively with respect to the electric field and spatial gradients of spins.
We denote the distribution function of order $a$ in the electric field and order $b$ in the gradient expansion by $f^a_b$.
Here, we consider contributions up to first order in the electric field and second order in spatial gradients.
Expanding the equilibrium distribution function $f_{\mathrm{eq}}$ in spatial gradients, we obtain
\begin{equation}
\begin{aligned}
    &f_\mathrm{eq} = f^0_0 + f^0_1 +  f^0_2 + \mathcal{O}(\lambda_s^{-3}), \\
    &f^0_0 = f_0,~
    f^0_1 = 0,~
    f^0_2 = ( - \bar{G}^{ij} \pi_i \pi_j + V   ) f_0'.
\end{aligned}
\end{equation}
Here, $f_0 = 1/(1 + e^{\beta( E(\bm{\pi}) - \mu)})~(E(\bm{\pi}) = \bm{\pi}^2/2m)$ is the unperturbed equilibrium distribution function.
Solving the Boltzmann equation, the perturbed distribution functions are given by
\begin{equation}
    \begin{aligned}
        &f^1_0 = \tau e E_i \partial_{\pi_i} f_0,~
        f^1_1 = 0, \\
        &f^1_2 = \tau e E_i \partial_{\pi_i} f^0_2 - \tau F_{ij} \frac{\pi_j}{m} \partial_{\pi_i} f^1_0.
    \end{aligned}
\end{equation}
The current is defined by
\begin{equation}
    j_i = -e \int_{\bm{x},\bm{\pi}} \dot{x}_i f.
\end{equation}
Here, we define $\int_{\bm{x},\bm{\pi}} \equiv  \int d^d\bm{x} /V \int d^d\bm{\pi}/(2\pi)^d $.
Using the distribution function linear in the electric field, the linear electrical conductivity is obtained as
\begin{equation}
\begin{aligned}
    &\sigma_{ij} = \sigma_{ij}^{\mathrm{L},0} + \sigma_{ij}^{\mathrm{L},G} + \sigma_{ij}^{\mathrm{H}}, \\
    &\sigma_{ij}^{\mathrm{L},0} = \frac{\tau e^2 n \delta_{ij}}{m} \\
    &\sigma_{ij}^{\mathrm{L},G} = - \frac{ \tau e^2 n }{m^2 J} \Bigl( \braket{G_{ij}} -\frac{\delta_{ij} }{2} \mathrm{tr} \braket{G} \Bigr) \\
    &~~~~~~~~~- \frac{\tau e^2 n' \mathrm{tr} \braket{G}}{2m^2} \delta_{ij}, \\
    &\sigma^\mathrm{H}_{ij} = \frac{\tau^2 e^2 n}{m^2} \braket{F_{ij}}.
\end{aligned}
\end{equation}
Here, $n = \int_{\bm{\pi}} f_0$ is the electron density and $n' = \partial n / \partial \mu$.
We use the notation $\braket{O} = \int d^d\bm{x} O/V$ for the spatial average.
$\sigma_{ij}^{\mathrm{L},0}$ is the conventional longitudinal electrical conductivity.
$\sigma_{ij}^{\mathrm{L},G}$ is the correction of the longitudinal conductivity by the Riemannian metric.
$\sigma_{ij}^{\mathrm{H}}$ is the transverse conductivity, which is known as the topological Hall conductivity.

\subsection{Nonreciprocal response induced by the geodesic SSC} \label{appendix_nonreciprocal_geodesic}
We consider a Hamiltonian corrected by the geodesic SSC $\Gamma_{ijk}$
\begin{equation}
    H = \frac{\bm{\pi}^2}{2m} - \bar{\Gamma}_{(ijk)} \pi_i \pi_j \pi_k,
\end{equation}
where we renormalize the coefficient such that $\bar{\Gamma}_{ijk} = \Gamma_{ijk}/ 4m^3 J^2$.
The parentheses $(ijk)$ indicate full symmetrization over the indices.
The canonical equations given by this Hamiltonian are
\begin{equation}
    \begin{aligned}
        \dot{x}_i &= \frac{\pi_i}{m} - 3 \bar{\Gamma}_{(ijk)} \pi_j \pi_k \\
        \dot{\pi}_i &= F_{ij} \dot{x}_j - e E_i + \mathcal{O}(\lambda^{-4}_s).
    \end{aligned}
\end{equation}
Expanding the equilibrium distribution function $f_{\mathrm{eq}}$ in spatial gradients, we obtain
\begin{equation}
\begin{aligned}
    &f_{\mathrm{eq}}  = f^0_0 + f^0_3 + \mathcal{O}(\lambda^{-4}_s), \\
    &f_0^0 = f_0,~f^0_1 = f^0_2 = 0,~f^0_3 = - \bar{\Gamma}_{(ijk)} \pi_i \pi_j \pi_k f'_0.\\
\end{aligned}
\end{equation}

In the following, we solve the Boltzmann equation up to second order in the electric field and third order in the gradient expansion.
The terms of second order in the electric field are given by
\begin{equation}
\begin{aligned}
    &f^2_0 = (\tau e)^2 E_a E_b \partial_{\pi_a \pi_b} f_0,~
    f^2_1 = f^2_2 =0,\\
    &f^2_3 = - (\tau e)^2 E_a E_b \bar{\Gamma}_{(ijk)} \partial_{\pi_a \pi_b} ( \pi_i \pi_j \pi_k f'_0).
    \end{aligned}
\end{equation}
Thus, the nonreciprocal conductivity reads
\begin{equation}
    \sigma_{ijk} = \frac{3 \tau^2 e^3 n}{2m^3J^2} \braket{\Gamma_{(ijk)}}.
\end{equation}

\bibliography{reference.bib}

\end{document}